\documentstyle[epsf]{mn}

\def\lapprox{\hbox{\lower .8ex\hbox{$\,\buildrel < \over\sim\,$}}}
\def\gapprox{\hbox{\lower .8ex\hbox{$\,\buildrel > \over\sim\,$}}}

\def\aj{AJ}
\def\apj{ApJ}
\def\aap{A\&A}

\def\aaps{A\&AS}

\def\mnras{MNRAS}

\def\nat{Nature}
\def\pasp{PASP}

\title{Monte Carlo simulations of the disk white dwarf population}

\author[E. Garc\'\i a-Berro et al.]
       {Enrique Garc\'\i a-Berro$^1$,
        Santiago Torres$^2$, 
        Jordi Isern$^3$, and
        Andreas Burkert$^4$\\
        $^1$Departament de F\'\i sica Aplicada, Universitat
	    Polit\`ecnica de Catalunya \& Institut d'Estudis 
	    Espacials de Catalunya/UPC, \\
	    Jordi Girona Salgado s/n, M\`odul B-4, Campus Nord, 
	    08034 Barcelona, Spain\\
        $^2$Departament de Telecomunicaci\'o i Arquitectura
	    de Computadors, E.U.P. de Matar\'o, Universitat
	    Polit\`ecnica de Catalunya, \\
	    Av. Puig i Cadafalch 101,
	    08303 Matar\'o, Spain\\
        $^3$Institut d'Estudis Espacials de Catalunya/C.S.I.C.,
	    Edifici Nexus, Gran Capit\`a 2-4, 08034 Barcelona, 
	    Spain\\
        $^4$Max-Planck-Institut f\"ur Astronomie, Koenigstuhl 17,
	    69117 Heidelberg, Germany}

\begin{document}

\maketitle

\begin{abstract}

In order to understand the dynamical and chemical evolution 
of our Galaxy it is of fundamental importance to study the 
local neighborhood. White dwarf stars are ideal candidates 
to probe the history of the solar neighborhood, since these 
``fossil'' stars have very long evolutionary time-scales and,
at the same time, their evolution is relatively well understood. 
In fact, the white dwarf luminosity function has been used for 
this purpose by several authors. However, a long standing problem 
arises from the relatively poor statistics of the samples, especially 
at low luminosities. In this paper we assess the statistical 
reliability of the white dwarf luminosity function by using 
a Monte Carlo approach. 

\end{abstract}


\section{Introduction}

The white dwarf luminosity function has become an important tool
to determine some properties of the local neighborhood, such as 
its age (Winget et al. 1987; Garc\'\i a-Berro et al. 1988; Hernanz
et al. 1994), or the past history of the star formation rate (Noh 
\& Scalo 1990; D\'\i az-Pinto et al. 1994; Isern et al. 1995a,b). 
This has been possible because now we have improved observational 
luminosity functions (Liebert, Dahn \& Monet 1988; Oswalt et al. 
1996; Leggett, Ruiz \& Bergeron 1998) and because we have a better 
understanding of the physics of white dwarfs and, consequently, 
reliable cooling sequences --- at least up to moderately low 
luminosities. 

The most important features of the luminosity function of white 
dwarfs are a smooth increase up to luminosities of $\log(L/L_{\sun})
\sim -4.0$, and the presence of a pronounced cut-off at $\log(L/
L_{\sun})\sim -4.4$, although its exact position is still today 
somehow uncertain since it hinges on the statistical significance 
of a small subset of objects, on how the available data is binned
and on the fine details of the sampling procedure. Most of the 
information on the early times of the past history of the local 
neighborhood is concentrated on this uncertain low luminosity portion 
of the white dwarf luminosity function. 

A major drawback of the luminosity function of white dwarfs is that 
it measures the volumetric density of white dwarfs and, therefore, 
in order to compare with the observations one must use the volumetric 
star formation rate, that is the star formation rate per cubic parsec, 
whereas for many studies of galactic evolution the star formation 
rate per square parsec is required and, consequently, fitted to 
the observations.

Another important issue is the fact that the sample from which the 
low luminosity portion $(M_{\rm V}>13^{\rm mag})$ of the white dwarf 
luminosity function is derived has been selected on a kinematical 
basis (white dwarfs with relatively high proper motions). Therefore, 
some kinematical biases or distortions are expected. Although there 
are some studies of the kinematical properties of white dwarf stars 
--- see, for instance, Sion et al. (1988) and references therein ---  
a complete and comprehensive kinematical study of the sample used 
to obtain the white dwarf luminosity function remains to be done. 
It is important to realize that a conventional approach to compute 
theoretical luminosity functions (Hernanz et al. 1994; Wood 
1992) does not take into account the kinematical properties 
of the observed sample. A Monte Carlo simulation of a model 
population of white dwarfs is expected to allow the biases 
and effects of sample selection to be taken into account, 
so their luminosity function could be corrected --- or, at 
least, correctly interpreted --- provided that a detailed 
simulation from the stage of source selection is performed 
accurately. Of course, a realistic model of the evolution 
of our Galaxy is required for that purpose.

Finally, the available white dwarf luminosity functions (Liebert
et al. 1988; Oswalt et al. 1996, Leggett et al. 1988) have been 
obtained using the $1/V_{\rm max}$ method (Schmidt 1968), which 
assumes a uniform distribution of the objects, yet nothing in our 
local neighborhood is, strictly speaking, homogeneous. In fact, 
stars in the solar neighborhood are concentrated in the plane of 
the galactic disk. Moreover, it is expected that old objects should 
have larger scale heights than young ones. This dependence on the 
scale height probably has effects on the observed white dwarf 
luminosity function --- especially on its low luminosity portion 
where old objects concentrate --- and, once again, a realistic model 
of galactic evolution is required for evaluating the effects of the 
departures from homogeneity of the observed samples. To our knowledge, 
this effect was only taken into account for the bright portion of the 
white dwarf luminosity function (Fleming, Liebert \& Green 1986) 
and not for the low luminosity portion where the effects are 
expected to be more dramatic. 

Perhaps the most sucessful application of the white dwarf luminosity 
function has been its invaluable contribution as an independent 
galactic chronometer to a better understanding of our Galaxy. 
Despite this fact there have been very few attempts --- being 
those of Garc\'\i a-Berro \& Torres (1997), Wood (1997) and Wood 
\& Oswalt (1998) the only serious ones --- to systematically 
investigate the statistical uncertainties associated with the 
derived age of the disk. Nevertheless the approach used by Wood 
\& Oswalt (1998) makes use of the {\sl observed} kinematic properties 
of the white dwarf population instead of using a standard model 
of the evolution of our Galaxy, which presumably should include 
the effects of a scale height law. Besides, these authors use the 
theoretical white dwarf luminosity function obtained from standard 
methods to assign probabilities and, ultimately, to assign 
luminosities to the white dwarfs of the sample. Finally, in 
their calculations Wood \& Oswalt (1998) computed the cooling 
times of all the white dwarfs of their sample by interpolating 
in a model cooling sequence of a 0.6~$M_{\sun}$ white dwarf, thus 
neglecting the effects of the full mass spectrum of white dwarfs. 

In this paper we explore the statistical reliability and 
completeness of the white dwarf luminosity function taking 
into account all of the above mentioned effects that were 
disregarded in previous studies. Special emphasis will be 
placed on the statistical significance of the reported cut-off 
in the white dwarf luminosity function. For that purpose we 
will use a Monte Carlo method, coupled with bayesian inference 
techniques, within the frame of a consistent model of galactic 
evolution, and using improved cooling sequences. 

To be precise, we want specific answers for the following questions: 
are the kinematics of the derived white dwarf population consistent 
with the observational data? Which are the effects of a scale height 
in the observed samples? Is the sample used to derive the white dwarf 
luminosity function representative of the whole white dwarf population? 
Or, at least, is this sample compatible with the white dwarf population 
within the limits imposed by the selection procedure? Which are the 
statistical errors for each luminosity bin? Which is the typical 
sampling error in the derived age of the disk?  

The paper is organized as follows: in \S 2 we describe how the 
simulated population of white dwarfs is built; in \S 3 we 
describe the kinematical properties of the samples obtained
in this way and we compare them with those of a real, although 
very preliminary and possibly uncomplete, sample; in \S 4 we 
study the spatial distribution of the samples, we assess the 
statistical reliability and completeness of the white dwarf 
luminosity function and we derive an estimate of the error 
budget in the determination of the age of the disk; finally, 
in \S 5 our results are summarized, followed by conclusions 
and suggestions for future improvements.

\section{Building the sample}

The basic ingredient of any Monte Carlo code is a generator of 
random variables distributed according to a given probability 
density. The simulations described in this paper have been done 
using a random number generator algorithm (James 1990) which 
provides a uniform probability density within the interval $(0,1)$ 
and ensures a repetition period of $\gapprox 10^{18}$, which is 
virtually infinite for practical simulations. When gaussian 
probability functions are needed we have used the Box-Muller 
algorithm as described in Press et al. (1986).

We randomly choose two numbers for the galactocentric polar coordinates
$(r,\theta)$ of each star in the sample within approximately 200 pc 
from the sun, assuming a constant surface density. The density changes 
due to the radial scale length of our Galaxy are negligible over the 
distances we are going to consider here and can be completely ignored. 
Next we draw two more pseudo-random numbers: the first for the mass 
$(M)$ on the main sequence of each star --- according to the initial 
mass function of Scalo (1998) --- and the second for the time at which 
each star was born $(t_{\rm b})$ --- according to a given star 
formation rate. We have chosen an exponentially drecreasing star 
formation rate per unit time and unit surface: $\psi\propto {\rm 
e}^{-t/\tau_{\rm s}}$. This choice of the shape of the star formation 
rate is fully consistent with our current understanding of the chemical 
evolution of our Galaxy --- see, for instance, Bravo et al. (1993). 
Once we know the time at which each star was born we assign the $z$ 
coordinate by drawing another random number according to an exponential 
disk profile. The scale height of newly formed stars adopted here 
decreases exponentially with time: $H_{\rm p}(t)=z_{\rm i}\,{\rm e}
^{-t/\tau_{\rm h}} + z_{\rm f}$. This choice for the time dependence 
of the scale height is essentially arbitrary, although it can be 
considered natural. We will however show that using these prescriptions 
for both the surface star formation rate and the scale height law to 
compute the theoretical white dwarf luminosity function leads to an 
excellent fit to the observations (see Isern et al. 1995a,b and \S 4) 
which does not result in a conflict with the observed kinematics of 
the white dwarf population (see \S 3). The values of the free 
parameters for both the surface star formation rate and the scale 
height have been taken from Isern et al. (1995a,b), namely: 
$\tau_{\rm s}=24$ Gyr, $\tau_{\rm h}=0.7$ Gyr, $z_{\rm i}/z_{\rm f}
=485$. 

In order to determine the heliocentric velocities in the $B3$ system, 
$(U,V,W)$, of each star in the sample three more quantities are drawn 
according to normal laws:
\begin{eqnarray}
n(U)&\propto&{\rm e}^{-(U-U'_0)^2/\sigma^2_{\rm U}}\nonumber\\
n(V)&\propto&{\rm e}^{-(V-V'_0)^2/\sigma^2_{\rm V}}\\
n(W)&\propto&{\rm e}^{-(W-W'_0)^2/\sigma^2_{\rm W}}\nonumber
\end{eqnarray}
where $(U'_0,V'_0,W'_0)$ take into account the differential rotation of 
the disk (Ogorodnikov 1965), and derive from the peculiar velocity 
$(U_{\sun},V_{\sun},W_{\sun})$ of the sun for which we have adopted 
the value $(10,5,7)\; {\rm km\; s^{-1}}$ (Dehnen \& Binney
1997).  

The three velocity dispersions $(\sigma_{\rm U},\sigma_{\rm V},
\sigma_{\rm W})$, and the lag velocity, $V_0$, of a given sample of 
stars are not independent of the scale height. From main sequence 
star counts, Mihalas \& Binney (1981) obtain the following relations, 
when the velocities are expressed in km~s$^{-1}$ and the scale height 
is expressed in kpc:
\begin{eqnarray}
U_0&=&0\nonumber\\
V_0&=&-\sigma^2_{\rm U}/120\nonumber\\
W_0&=&0\nonumber\\
&&\\
\sigma^2_{\rm V}/\sigma^2_{\rm U}&=&0.32+1.67\ 10^{-5}\sigma^2_{\rm U}
\nonumber\\
\sigma^2_{\rm W}/\sigma^2_{\rm U}&=&0.50\nonumber\\
H_{\rm p}&=&6.52\ 10^{-4}\sigma^2_{\rm W}\nonumber
\end{eqnarray}
which is what we adopt here (see as well \S 3.3). Note, however, that 
our most important input is the scale height law, from which most of 
the kinematical quantities are derived.

Since white dwarfs are long lived objects the effects of the 
galactic potential on their motion, and therefore on their 
positions and proper motions, can be potentially large, especially 
for very old objects which populate the tail of the white dwarf 
luminosity function. Therefore, the $z$ coordinate is integrated 
using the galactic potential proposed by Flynn et al. (1996). 
This galactic potential includes the contributions of the disk,
the bulge and the halo, and reproduces very well the local
disk surface density of matter and the rotation curve of our
Galaxy. We do not consider the effects of the galactic potential 
in the $r$ and $\theta$ coordinates. This is the same as assuming 
that the number of white dwarfs that enter into the sector of 
the disk that we are considering (the local column) is, on 
average, equal to the number of white dwarfs that are leaving 
it. Of course, with this approach we are neglecting the possibility
of a global radial flow, and thus, the possible effects of 
diffusion across the disk. However, the observed disk kinematics
suggest that radial mixing is efficient up to distances much larger 
than the maximum distance we have used in our simulations (Carney,
Latham \& Laird 1990). 

From this set of data we can now compute parallaxes and proper 
motions for all the stars $(\sim\,200\,000)$ in the sample.
Given the age of the disk $(t_{\rm disk})$ we can also compute 
how many of these stars have had time to evolve to white dwarfs 
and, given a set of cooling sequences (Salaris et al. 1997, 
Garc\'\i a--Berro et al. 1997), what are their luminosities. 
This set of cooling sequences includes the effects of phase 
separation of carbon and oxygen upon crystallization and has 
been computed taking into account detailed chemical profiles 
of the carbon-oxygen binary mixture present in most white dwarf 
interiors. These chemical profiles have been obtained using the 
most up to date treatment of the effects of an enhanced reaction 
rate for the $^{12}{\rm C}(\alpha,\gamma)^{16}{\rm O}$ reaction. 
Of course, a relationship between the mass on the main sequence 
and the mass of the resulting white dwarf is needed. Main sequence 
lifetimes must be provided as well. For these two relationships we 
have used those of Iben \& Laughlin (1989). The size of this new 
sample of white dwarfs typically is of $\sim\,60\,000$ stars 
(hereinafter ``original'' sample). Finally, for all white 
dwarfs belonging to this sample bolometric corrections are 
calculated by interpolating in the atmospheric tables of 
Bergeron et al. (1995) and their $V$ magnitude is obtained, 
assuming that all are non-DA white dwarfs.

\begin{figure}
\centering
\vspace*{12cm}
\includegraphics{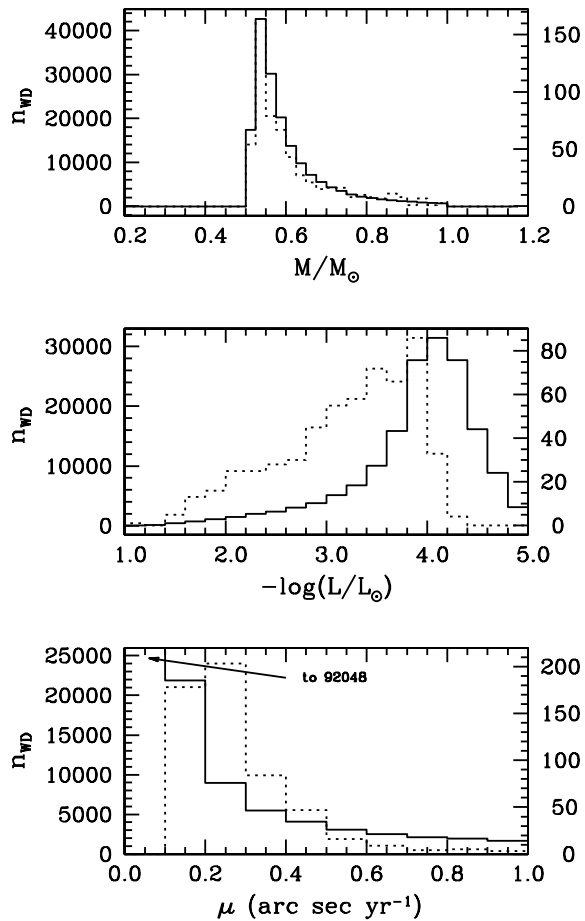}
\caption{Some relevant distributions obtained from a single Monte 
Carlo simulation, the solid lines correspond to the original sample 
whereas the dotted lines correspond to the restricted sample. See 
text for details.}
\end{figure}

Since the final goal is to compute the white dwarf luminosity 
function using the $1/V_{\rm max}$ method (Schmidt 1968) a set
of restrictions is needed for selecting a subset of white dwarfs
which, in principle, should be representative of the whole white
dwarf population. We have chosen the following criteria for 
selecting the final sample: $m_{\rm V}\le 18.5^{\rm mag}$ and 
$\mu\ge 0.16^{\prime\prime}\;{\rm yr}^{-1}$ (Oswalt et al. 1996). 
We do not consider white dwarfs with very small parallaxes $(\pi\le
0.005^{\prime\prime})$, since these are unlikely to belong to a 
realistic observational sample. All white dwarfs brighter than
$M_{\rm V}\le 13^{\rm mag}$ are included in the sample, regardless
of their proper motions, since the the luminosity function of hot
white dwarfs has been obtained from a catalog of spectroscopically 
identified white dwarfs (Green 1980; Fleming et al. 1986) which is 
assumed to be complete. Additionally all white dwarfs with tangential 
velocities larger than 250 km~s$^{-1}$ were discarded (Liebert, Dahn 
\& Monet 1989) since these would be probably classified as halo 
members. These restrictions determine the size of the the final 
sample which typically is $\sim\,200$ stars (hereinafter 
``restricted'' sample). Finally we normalize the total 
density of white dwarfs obtained in this way to its observed 
value in the solar neighborhood (Oswalt et al. 1996).

In Figure 1 we show a summary of the most relevant results for a 
disk age of 13 Gyr. In the top panel, the mass distribution of 
those stars that have been able to become white dwarfs (solid line, 
left scale) and of those white dwarfs that are selected for computing 
the luminosity function (dotted line, right scale) are shown. Both 
distributions are well behaved, follow closely each other, and peak 
at around $0.55\,M_{\sun}$ in very good agreement with the observations 
(Bergeron, Saffer \& Liebert 1992). In this sense, the restricted
sample could be considered as representative of the whole 
white dwarf population. 

In the middle panel of figure 1 we show the raw distribution of 
luminosities for the stars in the original (solid line, left scale) 
and the restricted (dotted line, right scale) samples. The 
differences between both distributions are quite apparent: first, 
the restricted sample has a broad peak centered at $\log(L/L_{\sun})
\sim -3.5$, whereas the original sample is narrowly peaked at a 
smaller luminosity (0.6 dex). Obviously, since the restricted
sample is selected on a kinematical basis --- see the lower 
panel of figure 1, where the distribution of proper motions 
for both samples is shown --- some very faint and low proper 
motion white dwarfs are discarded. Thus, the restricted sample 
is biased towards larger luminosities. Therefore, the cut-off 
of the observational luminosity function should be biased as 
well towards larger luminosities. However, it is important to 
realize that only $\sim$ 0.6\% of the total number of white 
dwarfs with $\log (L/L_{\sun})>-4.0$ are selected for the 
restricted sample, and therefore, used in computing the white 
dwarf luminosity function. This number decreases to $\sim$ 0.04\% 
if we consider the low luminosity portion of the white dwarf 
luminosity function --- that is, white dwarfs with $\log(L/L_{\sun})
<-4.0$ --- where most of the information regarding the initial 
phases of our Galaxy is recorded. The distribution of proper 
motions (lower right panel of figure 1) shows that most white 
dwarfs for both the original and the restricted sample have 
proper motions smaller than $0.4^{\prime\prime}$ yr$^{-1}$. 
However, the restricted sample has a pronounced peak at 
$\mu\sim\,0.3^{\prime\prime}$ yr$^{-1}$, and shows a deficit 
of very low proper motion white dwarfs, as should be the case 
for a kinematically selected sample, whereas the original 
sample smoothly decreases for increasing proper motions. 

\begin{figure*}
\centering
\vspace*{12cm}
\includegraphics{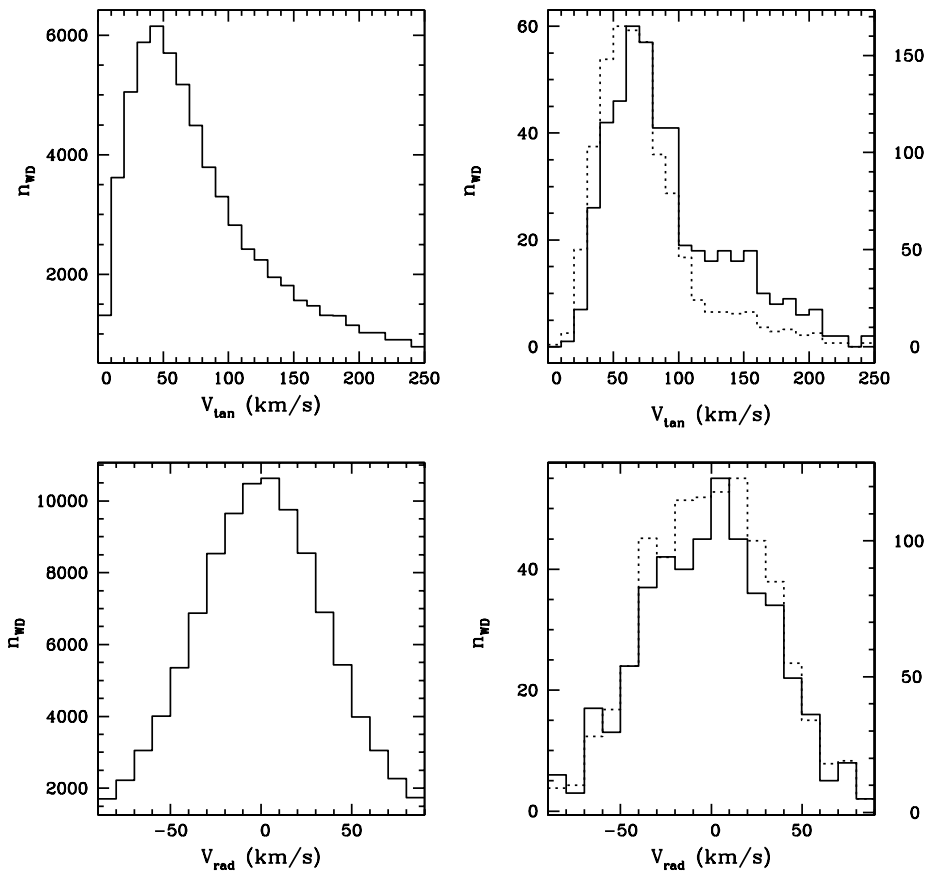}
\caption{Tangential (upper left panel) and radial (lower left panel) 
velocity distributions for the original sample and the corresponding 
distributions for the restricted sample (upper and lower right panels, 
respectively). Also shown as dotted lines are the tangential and 
radial velocity distributions of the restricted sample with a looser
restriction in proper motions (see text for details).}
\end{figure*}

\section{The kinematic properties of the white dwarf population}

Since the pioneering work of Sion \& Liebert (1977), very few 
analysis of the kinematics of the white dwarf population have 
been done, with that of Sion et al. (1988) being the most relevant 
one, despite the fact that the low luminosity portion of the 
white dwarf luminosity function is actually derived from a 
kinematically selected sample. Sion et al. (1988) used a specific 
subset of the proper-motion sample of spectroscopically identified 
white dwarfs to check kinematically distinct spectroscopic subgroups 
and test different scenarios of white dwarf production channels. 
However, a major disadvantage of this subset of the white dwarf 
population is that the three components of the velocity are derived 
only from the tangential velocity, since the determination of radial 
velocities for white dwarfs is not an easy task, especially for 
very cool ones. Obviously it would be better to have the complete 
description of the space motions of this sample, but it is nonetheless 
true that we already have two-thirds of the motion available for 
comparison with the simulated samples and that the latter samples
can account for this observational bias. 

The sample of Sion et al. (1988) consists of 626 stars with known 
distances and tangential velocities (of which 421 white dwarfs belong 
to the spectral type DA and 205 stars belong to other spectral types). 
In this proper motion sample there are 523 white dwarfs for which 
masses, radii and effective temperatures could be derived --- see 
Sion et al. (1988) for the computational details --- of which 372 
have masses larger than 0.5 $M_{\sun}$ and, therefore, are expected 
to have carbon-oxygen cores. Of this latter group of white dwarfs 
there are 305 with spectral type DA and 67 belong to other spectral 
types. For this particular sample of white dwarfs cooling ages were
derived using the cooling sequences of Salaris et al. (1997) and, 
given a relationship between the initial mass on the main sequence 
and the final mass of the white dwarf (Iben \& Laughlin 1989), main 
sequence lifetimes (Iben \& Laughlin 1989) were also assigned, and 
the birth time of their progenitors was computed. However, the
errors in the determination of the mass of the progenitor can 
produce large errors in the determination of the total age of
low mass white dwarfs. For instance, for a typical $0.6\,M_{\sun}$ 
white dwarf an error in the determination of its mass of 
$0.05\,M_{\sun}$ leads to an error in its cooling age of $\sim\,0.3$ 
Gyr at $\log(L/L_{\sun})=\,-2.0$ and of $\sim\,0.8$ Gyr at 
$\log(L/L_{\sun}) =\,-4.0$, whereas the error in the determination 
of its main sequence lifetime is of $\sim\,2$ Gyr. Thus, the mass 
dependence of the cooling sequences is relatively small, whereas 
the mass dependence of the main sequence lifetimes is very strong. 
Finally, it could be argued that since this sample includes both 
DA and non-DA white dwarfs, appropiate cooling sequences should be 
used for each spectral type. However, the errors introduced by using 
unappropiate cooling sequences (that is cooling sequences for 
He-dominated white dwarf envelopes) in the calculation of the 
cooling times of DA white dwarfs are small when compared to the 
errors introduced in dating white dwarfs by poor mass estimates. 
Therefore, the temporal characteristics of the white dwarf 
population from them derived should be viewed with some caution. 
Note as well that there is not any guaranty that the sample of Sion 
et al. (1988) is representative of the whole population of white 
dwarfs, since it is by no means complete, and therefore some cautions 
are required when drawing conclusions. To be more precise, the sample 
of Sion et al. (1988) has very few low luminosity white dwarfs. In 
fact, this sample contains only twelve white dwarfs belonging to the 
low luminosity sample of Liebert et al. (1988), of which only four 
have mass, tangential velocity, and effective temperature 
determinations. Therefore, we have added to this sample --- 
hereinafter ``observational'' sample --- three additional white 
dwarfs of the sample of Liebert et al. (1988) for which a mass 
estimate could be found (D\'\i az-Pinto et al. 1994). Nevertheless, 
this sample provides a unique opportunity to test the results 
obtained from a simulated white dwarf sample. 

\subsection{The overall kinematical properties of the samples}

First we compare the overall kinematical properties of the white
dwarf simulated samples with those of the observational sample, 
regardless of the birth time of their progenitors. In Figure 2 we 
show the distributions of the tangential and radial velocities for 
both the original and the restricted sample. The tangential velocity 
distribution of the original sample is shown in upper left panel of 
figure 2 and the tangential velocity distribution of the restricted 
sample is shown as a solid line (left scale) in the upper right panel. 
The restricted sample, which is kinematically selected, has a smaller 
tangential velocity dispersion $(\sigma_{\rm tan}\sim\,80$ km~s$^{-1})$ 
than the original sample $(\sigma_{\rm tan}\sim\,100$ km~s$^{-1})$. 
Here we have defined for operational purposes only the dispersions 
to be as the full width at half maximum of the distributions. Moreover, 
both samples are peaked at different tangential velocities: at 
$V_{\rm tan}\sim\,45$ km~s$^{-1}$ for the original sample and at 
$V_{\rm tan}\,\sim\,65$ km~s$^{-1}$ for the restricted sample, 
showing clearly that the restricted sample is biased towards 
larger tangential velocities, as it should be for a proper 
motion selected sample. In fact the most probable tangential 
velocity of the restricted sample is almost one third larger 
than that of the total sample of white dwarfs. This kinematical 
bias is clearly seen as well in the behavior of the distribution 
at low tangential velocities where the restricted sample shows a 
deficit of low velocity stars, as expected from a kinematically 
selected sample. Note as well the existence of an extended tail 
at high tangential velocities, indicating the presence of high 
proper motion white dwarfs. Of course, all these effects are 
simply due to the selection criteria and, in particular to the 
assumed restriction in proper motion. The distribution of radial 
velocities of the original sample is shown in the lower left panel 
of figure 2 and the radial velocity distribution of the restricted 
sample is shown as a solid line (left scale) of the lower right 
panel. Both distributions have similar dispersions $(\sigma_{\rm 
rad}\sim \,90$ km~s$^{-1}$) and both are well behaved and centered 
at $V_{\rm rad}=0$, as it should be since there is not any 
constrain on the radial velocities of the restricted sample.

\begin{figure}
\vspace{6.5cm}
\includegraphics{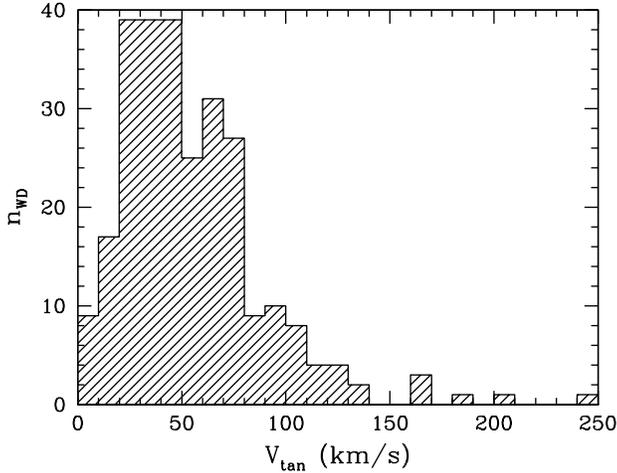}
\caption{Tangential velocity distribution of the sample of Sion et al. 
(1988).}
\end{figure}

In Figure 3 the tangential velocity distribution of the 
observational sample is shown. By comparing the tangential 
velocity distributions of figure 2 (upper panels) and figure 3 
we can assure that the observational sample does not have a clear 
kinematical bias since it does not show a clear deficit of low 
tangential velocity white dwarfs --- the ratio between the height 
of the peak and the height of the lowest velocity bin is the 
same for both the original sample and the observational sample: 
roughly 1/4 --- and does not have an extended tail at high tangential 
velocities as the restricted sample does. Moreover, the observational 
sample peaks at $V_{\rm tan}\sim\,40$ km~s$^{-1}$, whereas the original
sample (which is not kinematically selected) peaks at a very similar 
tangential velocity $(V_{\rm tan}\sim\,45$ km~s$^{-1})$. However the 
tangential velocity dispersion $(\sigma_{\tan}\sim\,60$ km~s$^{-1})$ 
of the observational sample is roughly one-third smaller than that 
of the original sample. This might be due to the absence of low 
luminosity white dwarfs in the observational sample. Notice that 
intrisically dim white dwarfs are selected on the basis of a large 
proper motion and, therefore, are expected to have, on average, larger 
tangential velocities, thus increasing the velocity dispersion. To 
check this assumption we have run our Monte Carlo code with a {\sl 
looser} restriction on proper motions $(\mu \ge 0.08^{\prime\prime}\;
{\rm yr}^{-1})$. The result is shown in the upper right panel of 
figure 2 as a dotted line (right scale). Although the number of 
selected white dwarfs increases from $\sim\,85$ to almost 250 the 
tangential velocity dispersion decreases from $\sigma_{\tan}\sim\,80$ 
km~s$^{-1}$ to $\sigma_{\tan}\sim\,60$ km~s$^{-1}$ in good agreement 
with the tangential velocity dispersion of the observational sample. 
A final test can be performed by imposing a {\sl tighter} restriction
on visual magnitudes ($m_{\rm V}\le 15.5^{\rm mag}$). The resulting
sample is now smaller --- 58 white dwarfs --- as should be expected,
whereas the tangential velocity dispersion decreases to $\sigma_{\tan}
\sim\,40$ km~s$^{-1}$ and the most probable tangential velocity remains
almost unchanged ($V_{\rm tan}\sim\,40$ km~s$^{-1}$). On the other hand 
the radial velocity distribution --- dashed line and right scale in the 
lower right panel of figure 2 --- is nearly indistinguishable from the 
previous sample, selected with a tighter restriction. Nevertheless, the 
differences between the observational sample and the simulated samples 
could be considered as minor. Therefore we conclude that the simulated 
population of white dwarfs is fairly representative of the real 
population of white dwarfs.

\subsection{The temporal behavior of the samples}

Up to this moment we have compared the global kinematical 
characteristics of the simulated samples with those of the 
sample of Sion et al. (1988), but one of the major advantages 
of this latter sample is that all the mass determinations 
have been obtained using the same procedure, and consequently, in 
this sense, the sample is relatively homogeneous. Therefore, we can 
tentatively obtain the temporal variations of the kinematical 
properties as a function of the birth time of the progenitors
of the white dwarfs belonging to the observational sample, and 
compare them with those of the simulated samples. 

\begin{figure}
\vspace{6.5cm}
\includegraphics{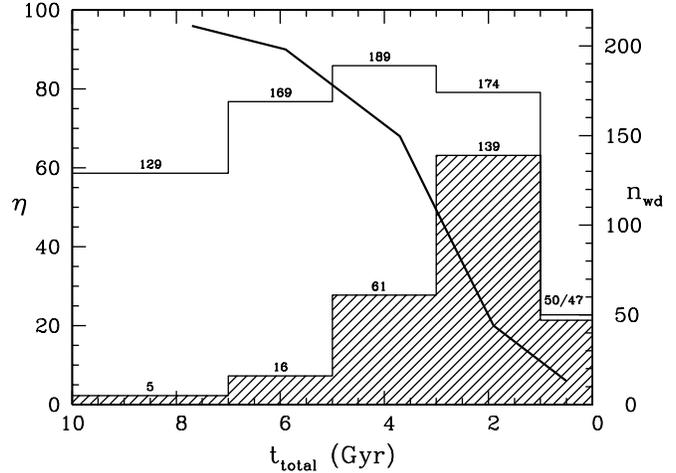}
\caption{Time distributions of the restricted sample (non-sahded 
diagram) and the observational sample (shaded diagram) and percentage 
of missing white dwarfs in the observational sample. The total number
of objects in each time bin is shown on top of the corresponding 
bin.}
\end{figure}

\begin{figure*}
\centering
\vspace*{16cm}
\includegraphics{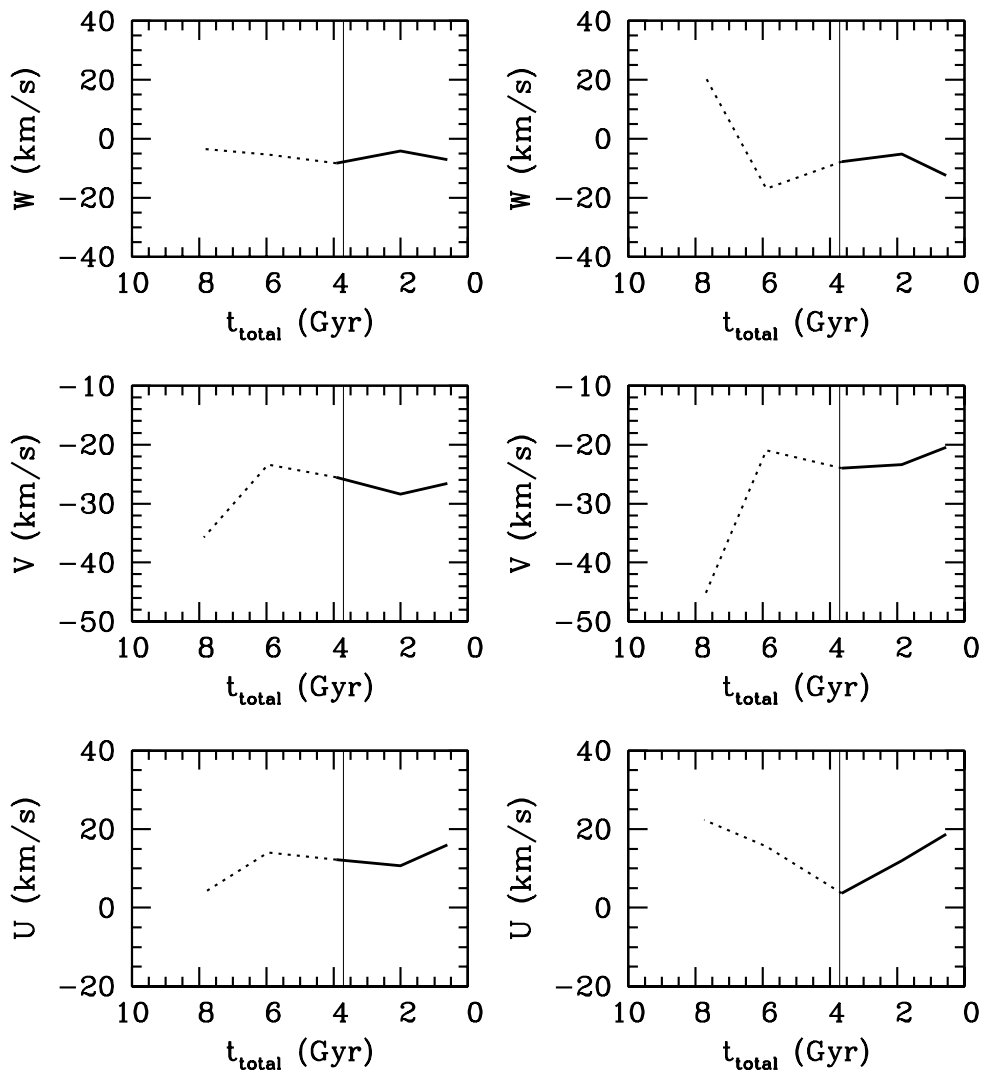}
\caption{Components of the tangential velocity as a function of the 
birth time of the white dwarf progenitors for the restricted (left-hand 
panels) and the observational sample (right-hand panels), see text 
for details.}
\end{figure*}

To this regard, in Figure 4 we show the histograms of the distribution 
of the birth times of white dwarfs belonging to the observational 
sample (shaded histogram) and to the restricted sample (non-shaded 
histogram). The number of objects in each time bin is also shown 
on top of each bin of the histogram. Time runs backwards and, 
therefore, old objects are located at the left of the diagrams, 
whereas young objects contribute to the time bins of the right 
part of the diagrams. It is important to realize that old bins 
may include kinematical data coming from either bright, low-mass 
white dwarfs, or dim, massive white dwarfs. The time bins have 
been chosen in such a way that the distribution of white dwarfs in 
the observational sample is efficiently binned. The first bin 
in time corresponds to objects older than 7~Gyr and has only 5 
objects, most of them corresponding to intrinsically faint objects 
belonging to the sample of Liebert et al. (1988). The last bin 
corresponds to objects younger than 1~Gyr. The remaining three 
bins are equally spaced in time and correspond to white dwarf 
progenitors with ages running from 1 to 7~Gyr in 2~Gyr intervals. 
All the bins have been centered at the average age of the objects 
belonging to them ($\sim$ 7.7, 5.9, 3.7, 1.9 and 0.5~Gyr, 
respectively). 

Since the youngest time bin corresponds to intrinsically bright white 
dwarfs it is expected that this time bin is reasonably complete in the 
observational sample. Therefore we have chosen the total number of 
stars in the simulated samples in such a way that the restricted sample 
has a number of objects in the youngest time bin comparable with that 
of the observational sample. Since there is no clear restriction in the 
ages of white dwarfs belonging to the restricted sample, the statistical
reliability of the remaining time bins of the observational sample 
can be readily assessed. Note the huge difference in the number of 
white dwarfs between the observational sample and the restricted 
sample for the oldest time bins of figure 4. Clearly, the completeness 
of the observational sample decreases dramatically as the birth time 
increases. The percentage of missing white dwarfs $(\eta)$ in the 
observational sample as a function of the birth time of their 
corresponding progenitors is also shown in figure 4 as a solid 
line, assuming that the youngest time bin of the restricted
sample is complete. We have considered the observational sample 
to provide reasonable estimates of the temporal variations 
of the velocity when one third of the expected number of white dwarfs 
is present in the corresponding time bin. This roughly corresponds 
to birth times smaller than 3.7~Gyr. Therefore, the only time bins
that we are going to consider statistically significant are the
youngest three bins.

In Figure 5 we show as solid lines the temporal variation of the 
components of the tangential velocity as a function of the total 
age (white dwarf cooling age plus main sequence lifetime of the 
corresponding parent star) of white dwarfs belonging to the 
restricted sample (left-hand panels), and the same quantities 
for white dwarfs belonging to the observational sample (right-hand 
panels). Although we have not considered the data for $t_{\rm total}
>3.7$ Gyr to be reliable due the incompleteness of the observational 
sample, we also show the temporal variations of all the three 
components of the tangential velocity for these times as dotted 
lines for the sake of completeness. The thinner vertical line 
corresponds to $t_{\rm total}=3.7$ Gyr. As can be seen in this 
figure, the general trend for young objects is very similar for 
both samples. In particular both the restricted and the observational 
sample have negative velocities across the galactic plane with 
velocities of $W\sim\,-10\,{\rm km~s}^{-1}$, both samples lag 
behind the sun with similar velocities of $V\sim\,-25\,{\rm km~s}
^{-1}$ and $-20\,{\rm km~s}^{-1}$, respectively, and both samples have 
positive radial velocities of roughly $U\sim 20\,{\rm km~s}^{-1}$. 
Finally, old objects in both samples lag behind the sun (middle 
panels) being the lag velocity comparable for both samples: 
$V\sim\,-20\,{\rm km~s}^{-1}$ and $-25\,{\rm km~s}^{-1}$, 
respectively. Moreover, we have computed the time-averaged 
values of the velocities shown in figure 5 and we have found 
$\langle U\rangle\sim\,10$ and $12$ km~s$^{-1}$, $\langle 
V\rangle\sim\,-28$ and $-23$ km~s$^{-1}$, and $\langle 
W\rangle\sim\,-8$ and $-7$ km~s$^{-1}$, respectively. We 
have computed as well the time-averaged values of the velocity 
dispersions for the restricted and the observational samples: 
$\langle\sigma_{\rm U} \rangle\sim\,41$ and 42 km~s$^{-1}$, 
$\langle\sigma_{\rm V} \rangle\sim\,27$ and 30 km~s$^{-1}$, 
and $\langle\sigma_{\rm W} \rangle\sim\,25$ and 25 km~s$^{-1}$, 
and we have found that they are also in good agreement. 

\begin{table}
\centering
\caption{Average values of the three components of the tangential
velocity and their corresponding dispersions (both in km~s$^{-1}$) for 
several choices of $z_{\rm f}$ (in kpc).}
\begin{tabular}{crcccrr}
\hline
$z_{\rm f}$ & 
$\langle U\rangle$  &
$\langle V\rangle$  &
$\langle W\rangle$  &
$\langle \sigma_{\rm U}\rangle$  &
$\langle \sigma_{\rm V}\rangle$  &
$\langle \sigma_{\rm W}\rangle$  \\
\hline
0.05 & 11.93 & $-13.87$ & $-3.54$ & 12.33 &  7.43 &  8.03 \\
0.10 & 12.77 & $-16.60$ & $-5.42$ & 20.42 & 11.09 & 10.85 \\
0.20 & 12.00 & $-19.86$ & $-4.02$ & 27.84 & 15.25 & 16.06 \\
0.30 &  9.40 & $-24.95$ & $-5.16$ & 31.52 & 19.54 & 20.70 \\
0.40 & 10.13 & $-24.93$ & $-7.33$ & 36.57 & 21.40 & 24.08 \\
0.50 &  9.34 & $-27.77$ & $-7.70$ & 41.38 & 26.75 & 24.74 \\
0.60 & 11.45 & $-29.90$ & $-5.83$ & 41.92 & 26.98 & 30.63 \\
\hline
\end{tabular}
\end{table}

As already noted in \S 2, the most important ingredient needed to fit 
adequately the kinematics of white dwarfs is the exact shape of the 
scale height law. In fact, the luminosity function (see section 4 
below) is only sensitive to the ratio of the initial to final scale 
heights $(z_{\rm i}/z_{\rm f})$ of the disk and to the time-scale of 
disk formation $(\tau_{\rm h})$ but not to the exact value of 
say $z_{\rm f}$. However when the kinematics of the sample are 
considered the reverse is true. That is the kinematics of the simulated
samples are very sensitive to the exact value adopted for the final 
scale height. This is clearly illustrated in Table 1, where the 
time-averaged values for the three components of the tangential 
velocity and the tangential velocity dispersions are shown for 
several choices of the final scale height, but keeping constant
the above mentioned ratio. As it can be seen there, the time-averaged 
radial component of the tangential velocity, $\langle U\rangle$, 
and the time-averaged perpendicular component of the tangential 
velocity, $\langle W\rangle$, are not very sensitive to the choice 
of $z_{\rm f}$, whereas the time-averaged lag velocity is very 
sensitive to its choice. Regarding the velocity dispersions all 
three components are sensitive. We have chosen the value of 
$z_{\rm f}$ which best fits the average values of the observed 
sample. In order to produce the results of figures 4 and 5 a 
value of 500 pc was adopted for $z_{\rm f}$, which is typical 
of a thick disk population. It is important to point out here 
that increasing (decreasing) $z_{\rm f}$ by a factor of two
without keeping constant the ratio $(z_{\rm i}/z_{\rm f})$
doubles (halves) $\sigma_{\rm W}$ for objects in the 
youngest time bin, which is the most reliable one, thus 
making incompatible the simulated and the observational 
samples. Similarly, increasing $\tau_{\rm h}$ by a factor 
of two changes dramatically the behaviour of the lag velocity 
since it changes the value of $V$ for objects in the youngest 
time bin from $\sim\,-20$ to $\sim\,-10$ km~s$^{-1}$. We conclude 
that the proposed scale height law is not in conflict with the 
observed kinematics of the white dwarf population.

\begin{figure}
\vspace{13cm}
\includegraphics{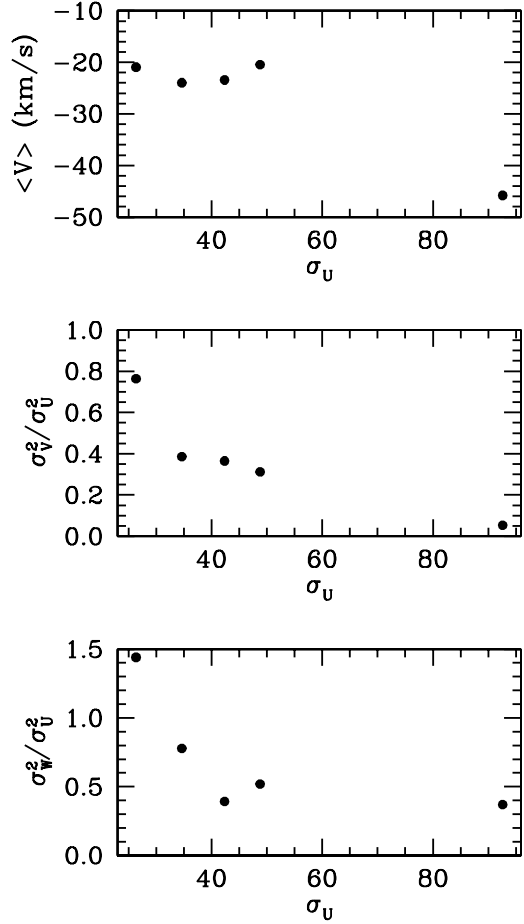}
\caption{Overall kinematical properties of the observational sample.}
\end{figure}

\subsection{A final remark on the reliability of the samples}

Finally, it is interesting to compare the results of a kinematical 
analysis of the observational sample with the predictions obtained 
from main sequence star counts as given by equation (2). In Figure 
6 we show the correlations between the $V$ component of the tangential 
velocity (top panel), the ratio $\sigma_{\rm V}/\sigma_{\rm U}$ 
(middle panel), and the ratio $\sigma_{\rm W}/\sigma_{\rm U}$ 
(bottom panel) as a function of the radial velocity dispersion 
$\sigma_{\rm U}$ obtained from the data of main sequence stars 
compiled by Sion et al. (1988). Except for the last bin the 
agreement between the data obtained from the white dwarf sample 
and the data obtained from main sequence stars is fairly good 
($\sigma_{\rm V}^2/\sigma_{\rm U}^2\sim\,0.3$, $\sigma_{\rm W}^2
/\sigma_{\rm U}^2\sim\,0.5$ and $V/\sigma_{\rm U}^2\sim\,-20$). 
However, it should be taken into account that the data coming 
from the last bin is obtained, as previously mentioned, with only 
five stars. Moreover, all these objects belong to the low luminosity 
sample of Liebert et al. (1988), which is strongly biased towards 
large tangential velocities and, besides, systematic errors affecting 
either mass and radius determinations or luminosity determinations 
(the bolometric corrections adopted in the latter work are highly 
uncertain) can mask the true behavior of the sample. 

\begin{table}
\centering
\caption{Average values of the three components of the tangential
velocity and their corresponding dispersions (both in km~s$^{-1}$) 
for the Monte Carlo simulation, the observational sample, and the 
Edvardsson et al. (1993) sample.}
\begin{tabular}{crcccrr}
\hline
Sample & 
$\langle U\rangle$  &
$\langle V\rangle$  &
$\langle W\rangle$  &
$\langle \sigma_{\rm U}\rangle$  &
$\langle \sigma_{\rm V}\rangle$  &
$\langle \sigma_{\rm W}\rangle$  \\
\hline
MC  & 10 & $ -28 $ & $ -8 $ & 41 & 27 & 25 \\
WD  & 12 & $ -23 $ & $ -7 $ & 42 & 30 & 25 \\
E93 & 14 & $ -21 $ & $ -8 $ & 39 & 29 & 23 \\
\hline
\end{tabular}
\end{table}

A final test of the validity of the assumptions adopted in this
paper to derive the simulated populations can be performed by
comparing the results of this section with the kinematical 
analysis of a sample of main sequence F and G stars (Edvardsson
et al. 1993). These authors measured distances, proper motions 
and radial velocities (among other data) for a sample of 189 F 
and G stars. They also assigned individual ages for all the stars 
in the sample from fits in the $T_{\rm eff}-\log g$ plane. The 
same sample has been re-analyzed very recently by Ng \& Bertelli 
(1998), using distances based on {\sl Hipparcos} parallaxes and 
improved isochrones. We refer the reader to the latter work for
a detailed analysis of the errors and uncertainties involved in 
dating individual objects. Although an analysis similar to that 
performed in \S 3.2 can be done, for the sake of conciseness 
we will only refer here to the average values of the three 
components of the tangential velocity and its corresponding 
dispersions. For this purpose in Table 2 we show the averaged 
values of the three components of the {\sl tangential velocity} and 
their corresponding dispersions for the restricted sample of our 
Monte Carlo simulation, labelled MC, the observational sample, 
labeled WD, and the three components of the {\sl velocity} and their
dispersions for the Edvardsson et al. (1993) sample, labeled E93. 
As already discussed in \S 3.2 the agreement between the Monte
Carlo simulation and the observational sample is fairly good. The 
comparison of both samples with the sample of Edvardsson et al. 
(1993) reveals that the agreement between the average values of 
the three samples is remarkably good, even if the dating procedure
for individual objects is very different in both observational
samples. The same holds for the averaged values of the three 
velocity dispersions. We conclude that our equation (2) represents 
fairly well the kinematical properties of the observed white dwarf 
population.

\section{The white dwarf luminosity function}

\subsection{The spatial distribution and completeness of the 
simulated white dwarf population}

The $1/V_{\rm max}$ method (Schmidt 1968, Felten 1976), when applied 
to our simulated white dwarf population, should provide us with an 
unbiased estimator of its luminosity function, pressumed completeness 
of the simulated samples in both proper motion and apparent magnitude, 
and provided that the spatial distribution of white dwarfs is 
homogeneous. Strictly speaking this means that the maximum distance 
at which we find an object belonging to the sample is independent of 
the direction. In our case this is clearly not true --- and, most 
probably, for a real sample this would certainly be the case as
well --- since we have derived the simulated samples assuming an 
exponential density profile across the galactic plane. Since the 
scale height law exponentially decreases with time (see \S 2) it 
is difficult to say ``a priori'' which is the final spatial 
configuration of the simulated white dwarf samples introduced 
in the previous sections.

\begin{figure}
\vspace{6.5cm}
\includegraphics{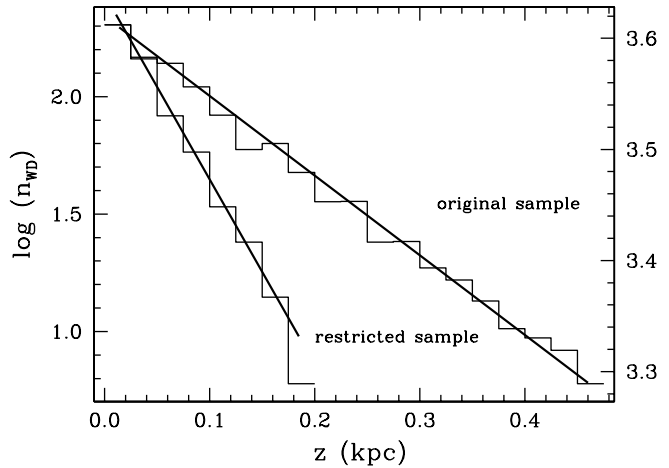}
\caption{Histogram of the $z$ distribution for both the original sample 
(right scale) and the restricted sample (left scale).}
\end{figure}

\begin{figure*}
\centering
\vspace*{12cm}
\includegraphics{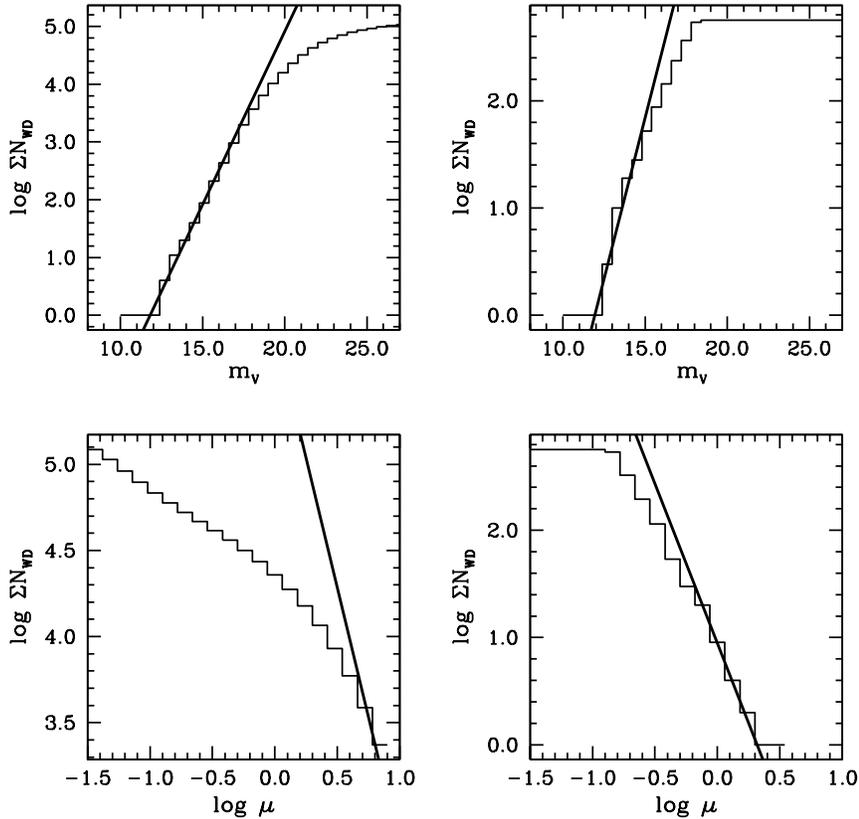}
\caption{Cumulative histograms of apparent magnitude and proper motion 
for both the original and the restricted sample. See text for details.}
\end{figure*}

In the histogram of Figure 7 we show the logarithmic distribution of 
the number of white dwarfs as a function of the absolute value of the 
$z$ coordinate for both the original sample (right scale) and the 
restricted sample (left scale). Clearly, both distributions correspond 
to exponential disk profiles with different scale heights. Also shown
in figure 7 are the best fits to these distributions. The corresponding
scale heights from them derived are $\approx 1.3$ kpc for the original 
sample, which is typical of a thick disk population, and considerably 
smaller $\approx 129$ pc for the restricted sample which can be 
considered typical of a thin disk population. This is not an evident 
result since, as has been explained in \S 2, the simulated populations 
take naturally into account the fact that old objects are distributed 
over larger volumes (that is, with larger scale heights and therefore 
with larger velocity dispersion perpendicular to the plane of the 
galaxy) than young ones. Therefore, one could expect that the final 
spatial distribution of the restricted white dwarf population --- 
which is kinematically selected --- should reflect properties of an 
intermediate thin-thick disk population and certainly this is not 
the case. Obviously, since there is not any restriction in the 
distances (within the local column) at which a white dwarf belonging 
to the original sample can be observed the expected final scale height 
for this sample should be much larger, in good agreement with the 
simulations. Regarding the restricted sample, our results clearly 
indicate that we are selecting for this sample white dwarfs lying 
very close to the galactic plane. Moreover, if we change by a factor 
of two $z_{\rm f}$ as explained in \S 3, the final scale height of the 
restricted sample does not change appreciably and, on the other hand, 
the dispersion of velocities perpendicular to the galactic plane does 
not agree with its observed value. Therefore, the final scale height 
of the restricted sample is clearly dominated by the selection 
criteria. It is important to realize that this scale height, taken 
at face value, is not negligible at all when compared with the value 
of the maximum distance at which a parallax is likely to be measured 
with relatively good accuracy --- which is typically 200 pc --- and 
which imposes an additional selection criterion (see \S 2) for white 
dwarfs belonging to the restricted sample, which are the white dwarfs 
which are going to be used in the process of determination of the white 
dwarf luminosity function. Therefore, the $1/V_{\rm max}$ method
must be generalized to take into account a space-density gradient. 
For this reason we have used the density law of figure 7 to define 
a new density-weighted volume element $dV'=\rho(z)\,dV$ (Felten
1976; Avni \& Bahcall 1980; Tinney, Reid \& Mould 1993), being $\rho
(z)$ the density law derived from figure 7. This new, corrected, 
estimator provides a more accurate determination of the white 
dwarf luminosity function and, ultimately, a more realistic 
value of the space density of white dwarfs. All in all, for 
reasonable choices of a scale height law its effects on the 
derived white dwarf luminosity function in principle cannot 
be considered negligible.

The second, and probably more important issue, is the completeness
of the samples used to build the white dwarf luminosity function.
This is a central issue since the $1/V_{\rm max}$ method assumes 
completeness of the samples. The reader should keep in mind that 
{\sl the original sample is complete by construction}, since it 
consists of all white dwarfs generated by the Monte Carlo code, 
regardless of their distance, proper motion, apparent magnitude 
and tangential velocity; whereas the restricted sample is built with 
white dwarfs culled from the original sample according to a set of 
selection criteria and, therefore, its completeness remains to be 
assessed. 

In Figure 8 we explore the completeness of the simulated 
samples. For this purpose, the cumulative star counts of white 
dwarfs with apparent magnitude smaller than $m_{\rm V}$ for the 
original sample are shown in the top left panel of figure 8, whereas
the corresponding diagram for the restricted sample is shown in the 
top right panel. Also shown in figure 8 are the cumulative star
counts of white dwarfs with proper motions larger than $\mu$
belonging to the original sample (bottom left panel) and to 
the restricted sample (bottom right panel). For a complete sample 
distributed according to a homogenous spatial density, the 
logarithm of the cumulative star counts of white dwarfs with 
apparent magnitude smaller than $m_{\rm V}$ are proportional 
to $m_{\rm V}$ with a slope of 0.6 (see, for instance, Mihalas
\& Binney 1981). We also show in the top panels of figure 8 a 
straight line with such a slope. It is evident from the previous 
discussion that our samples are not, by any means, distributed 
homogenously. Note as well that the effects of a scale height 
law are tangled in the standard test of completeness of the 
samples. Nevertheless, the effects of a scale height law can be
disentangled since they should be quite apparent in the 
cumulative star counts diagram of the original sample, which is 
complete. A look at the top left panel of figure 8 reveals 
that the effects of the scale height law are evident for 
surveys with limiting magnitude $m_{\rm V}\gapprox 19^{\rm mag}$. 
Therefore, we can now assess the completeness in apparent 
magnitude of the restricted sample, since the turn-off for this 
sample (see top right panel of figure 8) occurs at $m_{\rm
V}\sim\,17^{\rm mag}$. Consequently, the effects of the scale 
height law can be completely ruled out, and this value can be 
considered as a safe limit for which the restricted sample is 
complete in apparent magnitude.

The completeness of the restricted sample in proper motion can be 
assessed in a similar way. Again, the assumption of an homogenous
and complete sample in proper motion leads to the conclusion that
the logarithm of the cumulative star counts of white dwarfs with 
proper motion larger than $\mu$ should be proportional to $\mu$ 
with a slope of $-3$ (see, for instance, Oswalt \& Smith 1995, 
and Wood \& Oswalt 1998). A look at the bottom left panel of 
figure 8 reveals that for the original sample this is not by 
far the case. In other words, since this particular sample is 
complete by construction the hypothesis of an homogenous 
distribution of proper motions must be dropped. This is again 
one, and probably the most important, of the effects associated 
with a scale height law since the kinematics of the samples are
highly sensitive to the choice of the scale height law (see
\S 2, equation (2) and table 1). It is important to realize
that the effects of a scale height law are more prominent 
in proper motion than in the spatial distribution and this
can be directly checked for a real sample, thus providing 
a direct probe of the history of the star formation rate
per unit volume. Finally, in the lower right panel of 
figure 8 the cumulative star counts in proper motion of 
white dwarfs belonging to the restricted sample are shown.
As expected, the effects of a scale height law are in 
this case negligible since we are culling white dwarfs
with high proper motion for which the original sample is 
reasonably complete (see the lower left panel of figure
8). The exact value of the turn-off is in this case
$\mu\sim\, 0.3^{\prime\prime}\;{\rm yr}^{-1}$, in close
agreement with the results of Wood \& Oswalt (1998).
 
It is quite clear from the previous discussions that one 
of the ingredients that has proven to be essential in the 
determination of the white dwarf luminosity function is 
the adopted scale height law. In principle one should 
expect two kinds of competing trends. On the one hand, the 
effects of the scale height law should be more dramatic 
for old objects, because old objects have a larger velocity
dispersion (although the effects of a spatial inhomogeneity 
should be, as well, less apparent) and the tail of the white 
dwarf luminosity function is populated predominantly by this 
kind of white dwarfs (intrinsically dim, high proper motion
objects). Therefore, one should expect that the cut-off in 
the white dwarf luminosity function is influenced either by 
the spatial distribution of white dwarfs or by their velocity 
distribution or by a combination of both. On the other hand, 
objects populating the tail of the luminosity function are 
intrinsically dim objects and, therefore, in order to be 
selected for the restricted sample they must be close neighbors. 
This, in turn, implies that the average distance at which we are 
looking for white dwarfs is small and, consequently, the effects 
of a scale height law should be less apparent. The reverse is 
true at moderately high luminosities. Therefore it is interesting 
to see which are the dominant effects as a function of the luminosity. 
For this purpose, in Figure 9 we show several average properties of 
white dwarfs belonging to the restricted sample as a function of their 
luminosity for a typical Monte Carlo simulation. 

\begin{figure}
\vspace*{12cm}
\includegraphics{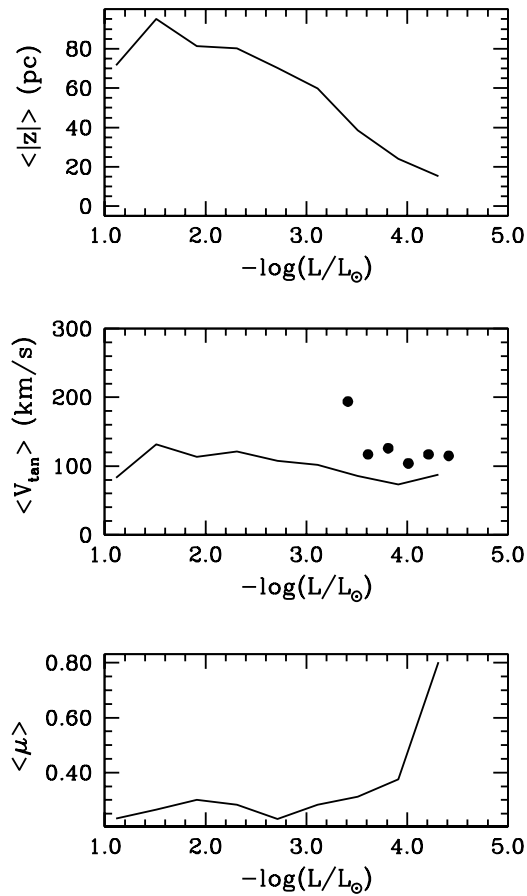}
\caption{Average properties of the restricted sample as a function of 
the luminosity. Bottom panel: average $z$ coordinate of white dwarfs of 
the restricted sample; middle panel: average tangential velocity for 
these white dwarfs, the observational data has been taken from Liebert 
et al. (1988); and top panel: average proper motion of these objects.}
\end{figure}

In the top panel of figure 9 we show the average distance to 
the galactic plane of white dwarfs belonging to the restricted sample. 
As it can be seen, the average distance to the galactic plane of 
intrinsically bright white dwarfs can be as high as 100 pc, 
which is a sizeable fraction of the derived scale heights of 
the white dwarf samples. Consequently we expect that the effects 
of an inhomogeneous spatial distribution should be very prominent 
at high luminosities. Conversely, the average distance to the galactic 
plane for white dwarfs near the observed cut-off in the white dwarf 
luminosity function is only $\sim\,10$ pc. Therefore as the luminosity 
decreases we are probing smaller volumes and the effects of an 
inhomogenous {\sl spatial} distribution at low luminosities are 
expected to be, from this point of view, small (but see, however, 
the discussion in \S 4.4). 

In the middle panel of figure 9 the average tangential velocity 
of objects belonging to the restricted sample is shown as a function 
of the luminosity. The observational data is shown as solid circles 
and has been obtained from Liebert et al. (1988). Their adopted 
restriction in proper motion $\mu_0=0.80^{\prime\prime}\;{\rm yr}
^{-1}$ is significantly larger than the one we adopt here $\mu_0= 
0.16^{\prime\prime}\;{\rm yr}^{-1}$, which is consistent with the 
cut-off in proper motion adopted by Oswalt et al. (1996). Therefore 
we expect a smaller average tangential velocity. The agreement is 
fairly good since, given the ratio of proper motion cut-offs, the 
average tangential velocity of our restricted sample should be roughly 
a 20\% smaller: a closer look at the middle panel of figure 9 
shows that the average tangential velocity reported by Liebert et 
al. (1988) is $\sim\,120$ km~s$^{-1}$, whereas we obtain $\sim\,90$ 
km~s$^{-1}$. These figures reinforce the general idea that our 
simulations are fully consistent with the observed kinematics of
the white dwarf population. 

Finally, in the bottom panel of figure 9 the average proper motion 
distribution of those stars belonging to the restricted sample is 
shown as a function of the luminosity. As it can be seen there, low 
luminosity white dwarfs belonging to the restricted sample have, on 
average, large proper motions. As expected from the discussion of the 
two previous panels, the distribution of proper motions is smoothly 
increasing for luminosities in excess of $\sim\,10^{-3}\,L_{\sun}$:
since the average value of the tangential velocity remains 
approximately constant, and we are selecting objects with smaller 
average distances, the net result is an increase in the average 
proper motion. Moreover, white dwarfs belonging the low luminosity 
portion of the white dwarf luminosity function are preferentially 
culled from the original sample because of their high proper motion. 
That is the same to say that the selection criterion is primarily 
the proper motion one and that the criterion on apparent magnitude 
has little to do for these luminosities, in agreement with the results 
of Wood \& Oswalt (1988). As a final consequence the effects of an 
inhomogenous distribution in {\sl proper motions} will be more 
evident at high luminosities, where the average proper motion 
is smaller (see the discussion of the lower left panel of figure
8). All in all, the effects of the inhomogeneities in both 
proper motion and $z$ will be more prominent at high luminosities,
where the observational luminosity function already takes into
account these effects (Fleming et al. 1986).

\begin{table*}
\centering
\caption{Total number of white dwarfs, $N_{\rm WD}$, and white 
dwarfs in each bin, $N_{\rm i}$, for each of the twenty realizations 
of the simulated white dwarf luminosity functions.}
\begin{tabular}{rrrrrrrrrrr}
\hline
$i$ & 
$N_{\rm WD}$ &
$N_1$ &
$N_2$ &
$N_3$ &
$N_4$ &
$N_5$ &
$N_6$ &
$N_7$ &
$N_8$ &
$N_9$ \\
\hline
   1& 200& 1&  8&  5& 14& 42&  38&  42&  44&  6 \\
   2& 216& 1&  6& 11& 18& 29&  48&  49&  53&  1 \\
   3& 203& 0&  4&  8& 22& 36&  39&  45&  43&  6 \\
   4& 176& 0&  6&  6& 17& 18&  35&  41&  44&  9 \\
   5& 210& 1&  5&  8& 17& 24&  49&  48&  53&  5 \\
   6& 191& 1&  7& 10& 23& 24&  37&  44&  37&  8 \\
   7& 202& 0&  3& 12& 29& 27&  42&  50&  35&  4 \\
   8& 222& 0&  1& 16& 18& 38&  41&  50&  57&  1 \\
   9& 197& 0&  5& 12& 20& 22&  34&  51&  47&  6 \\
  10& 198& 1&  1&  9& 16& 28&  44&  49&  47&  3 \\
  11& 204& 0&  5& 10& 21& 25&  38&  44&  53&  8 \\
  12& 198& 1&  3& 14& 20& 33&  36&  50&  35&  6 \\
  13& 175& 0&  3& 14& 20& 23&  32&  44&  37&  2 \\
  14& 182& 0&  6&  9& 15& 28&  40&  43&  35&  6 \\
  15& 185& 2&  6&  6& 19& 30&  31&  44&  43&  4 \\
  16& 213& 1&  3& 16& 21& 33&  31&  45&  56&  7 \\
  17& 189& 2&  2& 11& 25& 22&  32&  43&  44&  8 \\
  18& 207& 1&  4&  7& 17& 33&  38&  41&  61&  5 \\
  19& 217& 1& 10& 10& 28& 17&  47&  55&  42&  7 \\
  20& 210& 0&  4& 13& 19& 31&  38&  46&  55&  4 \\
\hline
\end{tabular}
\end{table*}
         
\begin{table*}
\centering
\caption{$\chi^2$ test of the compatibility of the Monte Carlo simulated
samples.}
\begin{tabular}{cccccccccccccccccccc}
\hline
$i$ & 
2 &
3 &
4 &
5 &
6 &
7 &
8 &
9 &
10 &
11 & 
12 & 
13 & 
14 & 
15 &
16 &
17 &
18 & 
19 & 
20  \\
\hline
1&        0.24&  0.80&  0.21&  0.41&  0.37&  0.05&  0.03&  
   0.16&  0.24&  0.33&  0.32&  0.04&  0.70&  0.86&  0.17&  
   0.06&  0.75&  0.02&  0.34 \\
2&            &  0.46&  0.09&  0.93&  0.26&  0.31&  0.59& 
   0.48&  0.79&  0.48&  0.30&  0.37&  0.36&  0.47&  0.29& 
   0.11&  0.67&  0.19&  0.86 \\
3&            &      &  0.40&  0.69&  0.81&  0.78&  0.22& 
   0.73&  0.69&  0.88&  0.90&  0.44&  0.89&  0.85&  0.67& 
   0.56&  0.81&  0.16&  0.89 \\
4&            &      &      &  0.63&  0.86&  0.13&  0.01& 
   0.81&  0.20&  0.88&  0.19&  0.24&  0.77&  0.60&  0.14& 
   0.57&  0.36&  0.37&  0.26 \\
5&            &      &      &      &  0.65&  0.35&  0.16& 
   0.80&  0.90&  0.92&  0.38&  0.26&  0.68&  0.65&  0.39& 
   0.43&  0.90&  0.54&  0.84 \\
6&            &      &      &      &      &  0.77&  0.01& 
   0.92&  0.32&  0.89&  0.86&  0.59&  0.94&  0.85&  0.44& 
   0.89&  0.32&  0.83&  0.49 \\
7&            &      &      &      &      &      &  0.11& 
   0.68&  0.52&  0.47&  0.89&  0.80&  0.55&  0.34&  0.27& 
   0.61&  0.11&  0.42&  0.50 \\
8&            &      &      &      &      &      &      & 
   0.12&  0.67&  0.11&  0.23&  0.20&  0.03&  0.06&  0.46& 
   0.04&  0.31&  0.01&  0.76 \\
9&            &      &      &      &      &      &      & 
       &  0.58&  0.99&  0.77&  0.82&  0.77&  0.68&  0.77& 
   0.83&  0.53&  0.59&  0.92 \\
10&           &      &      &      &      &      &      & 
       &      &  0.54&  0.69&  0.58&  0.54&  0.57&  0.45& 
   0.51&  0.74&  0.07&  0.81 \\
11&           &      &      &      &      &      &      & 
       &      &      &  0.56&  0.41&  0.72&  0.67&  0.85& 
   0.81&  0.87&  0.46&  0.95 \\
12&           &      &      &      &      &      &      & 
       &      &      &      &  0.79&  0.84&  0.67&  0.77& 
   0.77&  0.29&  0.15&  0.71 \\
13&           &      &      &      &      &      &      & 
       &      &      &      &      &  0.62&  0.51&  0.38& 
   0.60&  0.15&  0.10&  0.63 \\
14&           &      &      &      &      &      &      & 
       &      &      &      &      &      &  0.79&  0.24& 
   0.38&  0.40&  0.25&  0.56 \\
15&           &      &      &      &      &      &      & 
       &      &      &      &      &      &      &  0.47& 
   0.65&  0.82&  0.16&  0.61 \\
16&           &      &      &      &      &      &      & 
       &      &      &      &      &      &      &      & 
   0.78&  0.79&  0.04&  0.96 \\
17&           &      &      &      &      &      &      & 
       &      &      &      &      &      &      &      & 
       &  0.38&  0.28&  0.50 \\
18&           &      &      &      &      &      &      & 
       &      &      &      &      &      &      &      & 
       &      &  0.04&  0.93 \\
19&           &      &      &      &      &      &      & 
       &      &      &      &      &      &      &      & 
       &      &      &  0.12 \\
\hline
\end{tabular}
\end{table*}

\subsection{The Monte Carlo simulated white dwarf luminosity
functions}

\begin{figure}
\centering
\vspace{15cm}
\includegraphics{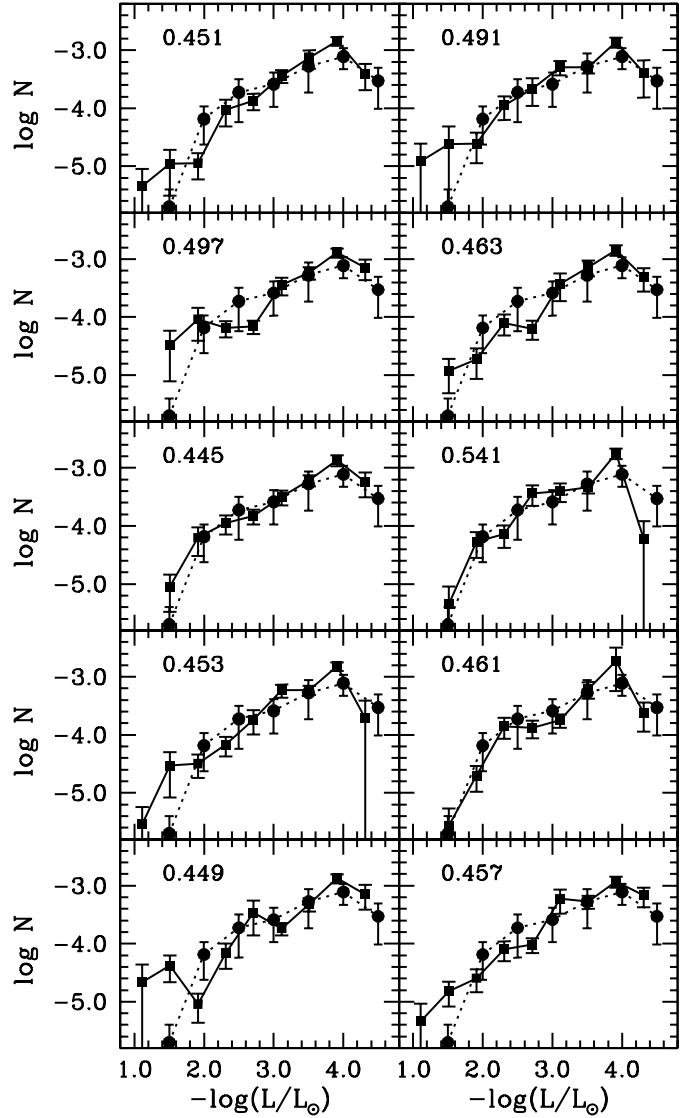}
\caption{Panel showing different realizations of the simulated white 
dwarf luminosity function --- filled squares and solid lines ---
compared to the observational luminosity function of Oswalt et al.
(1996) --- filled circles and dotted line.}
\end{figure}

In Figure 10 we show a set of panels containing the white dwarf 
luminosity functions obtained from ten different Monte Carlo 
simulations. That is the same to say that ten different initial 
seeds were chosen for the random number generator and, consequently, 
ten {\sl independent realizations} of the white dwarf luminosity 
function were computed (in fact, we have computed twenty independent 
realizations, of which only ten are shown in figure 10). The 
adopted age of the disk was $t_{\rm disk}=13$~Gyr and the set
of restrictions used to build the sample is that of \S 2, which
is the same set used by Oswalt et al. (1996) to derive their
observational white dwarf luminosity function. The simulated 
white dwarf luminosity functions were computed using a generalized 
$1/V_{\rm max}$ method (Felten 1976; Tinney et al. 1993; Qin \& Xie 
1997) which takes into account the effects of the scale height. The 
error bars of each bin were computed according to Liebert et al. 
(1988): the contribution of each star to the total error budget 
in its luminosity bin is conservatively estimated to be the same 
amount that contributes to the resulting density; the partial 
contributions of each star in the bin are squared and then added,
the final error is the square root of this value. The resulting 
white dwarf luminosity functions are plotted as solid squares; a 
solid line linking each one of their points is also shown as a 
visual help. Also plotted in each one of the panels is the 
observational white dwarf luminosity function of Oswalt et al. 
(1996), which is shown as solid circles linked by a dotted line. 
For each realization of the white dwarf luminosity function the 
value obtained for $\langle V/V_{\rm max} \rangle$ is also shown 
in the upper left corner of the corresponding panel. Finally, 
and for sake of completeness, the total number of objects in 
the restricted sample, $N_{\rm WD}$, of the different realizations 
of the white dwarf population, and the distribution of objects, 
$N_{\rm i}$, in each luminosity bin, $i$, of the Monte Carlo 
simulated white dwarf luminosity functions are shown in Table 
3. The total number of white dwarfs belonging to the restricted 
sample is roughly 200, which is the typical size of the samples 
used to build the currently available observational luminosity 
functions. This number is important since the assigned error 
bars are strongly dependent on the number of objects in each 
luminosity bin.

\begin{figure*}
\centering
\vspace*{16cm}
\includegraphics{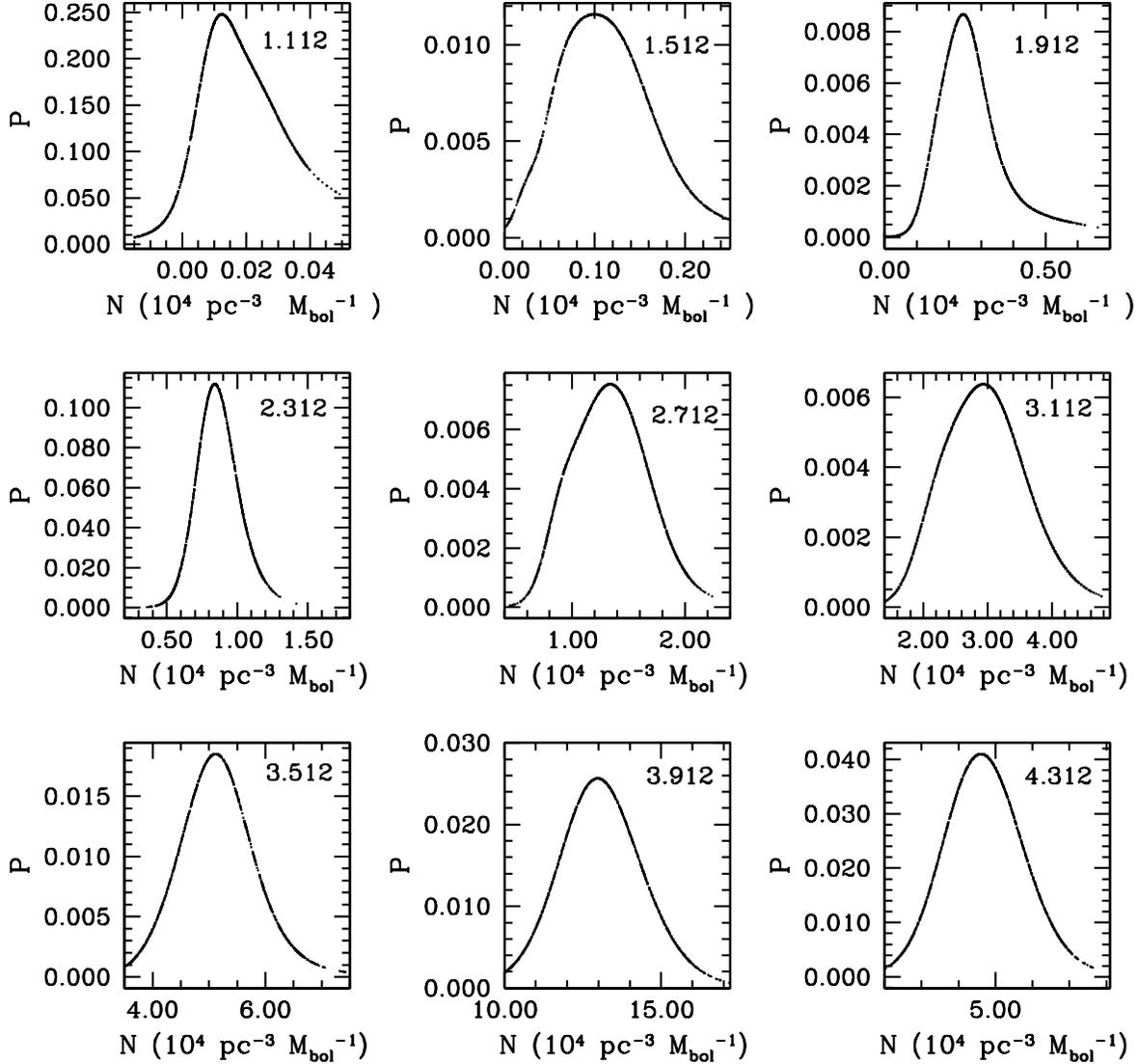}
\caption{Probability distribution functions for each luminosity bin.}
\end{figure*}
         
It is important to notice the overall excellent agreement between 
the simulated data and the observational luminosity function. 
However, there are several points that deserve further comments. 
The first one is that the simulated white dwarf luminosity functions
are systematically larger than the observational luminosity 
functions for luminosities in excess of $\log(L/L_{\sun})=-2.0$.
This behavior reflects the effects of the spatial inhomogeneity
ot the simulated white dwarf samples. It is important to realize
that the hot portion of the white dwarf luminosity function of 
Oswalt et al. (1996) has been derived without taking into account
the effects of a scale height, in contrast with the procedure
adopted by  Fleming et al. (1986), where those effects were 
properly taken into account. When one compares the luminosity 
functions obtained in this section with that of Fleming et al.
(1986) the agreement is excellent. Also of interest is the 
fact that the hot portion of the white dwarf luminosity 
function varies quite considerably for the different realizations. 
The reason for this behavior is that at high luminosities the 
evolution is dominated by neutrino losses and it is fast. Therefore, 
the probability of finding such white dwarfs is relatively small 
and the statistical signifance of those bins is low. Consequently, 
the exact shape of the luminosity function at $\log(L/L_{\sun}) 
\ge -3.0 $ is strongly dependent of the initial seed of the 
pseudo-random number generator. This is further confirmed by 
comparing the second and the third column of table 3 where the 
total number of objects in the restricted sample and the number of 
objects in the first bin of the white dwarf luminosity function
of each realization of the Monte Carlo simulations are shown.
As a consequence the real error bars that should be assigned to 
each bin are presumably larger than those of figure 10. Moreover, 
any attempt to derive the volumetric star formation rate using 
data from the bins at high luminosities (Noh \& Scalo, 1990) is 
based on very weak grounds. It is also important to notice that 
the completeness of the simulated samples as derived from the 
value of $\langle V/V_{\rm max} \rangle$ is relatively large. 
In fact, for a complete and homogeneous sample this value 
should be equal to 0.5; since the simulated sample samples are 
not homogenous the values obtained here can be considered as 
reasonable. 

\begin{table*}
\centering
\caption{Error bars of the twenty independent realizations of the
Monte Carlo simulated white dwarf luminosity functions in each
luminosity bin and the same quantities for the Bayesian luminosity
function (last row).}
\begin{tabular}{cccccccccc}
\hline
$i$ & 
$\Delta \log(n_1)$ &
$\Delta \log(n_2)$ &
$\Delta \log(n_3)$ &
$\Delta \log(n_4)$ &
$\Delta \log(n_5)$ &
$\Delta \log(n_6)$ &
$\Delta \log(n_7)$ &
$\Delta \log(n_8)$ &
$\Delta \log(n_9)$ \\
\hline
 1& 1.000&0.484&0.517&0.464&0.597&0.238&0.261&0.180&0.454 \\
 2& 1.000&0.718&0.432&0.368&0.446&0.253&0.179&0.163&1.000 \\
 3& 0.000&0.631&0.514&0.363&0.283&0.285&0.215&0.216&0.453 \\
 4& 0.000&0.877&0.537&0.524&0.393&0.272&0.224&0.190&0.525 \\
 5& 1.000&0.723&0.476&0.481&0.319&0.256&0.347&0.171&0.474 \\
 6& 1.000&0.461&0.431&0.376&0.285&0.447&0.251&0.210&0.372 \\
 7& 0.000&0.974&0.474&0.392&0.340&0.262&0.316&0.699&0.517 \\
 8& 0.000&1.000&0.473&0.417&0.385&0.353&0.213&0.206&1.000 \\
 9& 0.000&0.590&0.539&0.387&0.362&0.500&0.304&0.205&0.439 \\
10& 1.000&1.000&0.541&0.440&0.496&0.275&0.213&0.195&0.628 \\
11& 0.000&0.763&0.577&0.312&0.258&0.338&0.245&0.177&0.381 \\
12& 1.000&0.746&0.441&0.446&0.404&0.240&0.435&0.201&0.429 \\
13& 0.000&0.618&0.450&0.362&0.373&0.270&0.306&0.191&0.775 \\
14& 0.000&0.741&0.644&0.486&0.281&0.344&0.195&0.190&0.425 \\
15& 0.768&0.887&0.560&0.409&0.292&0.357&0.423&0.178&0.824 \\
16& 1.000&0.890&0.450&0.326&0.335&0.344&0.214&0.332&0.458 \\
17& 0.785&0.965&0.446&0.318&0.497&0.324&0.203&0.171&0.408 \\
18& 1.000&0.537&0.496&0.536&0.349&0.254&0.197&0.161&0.484 \\
19& 1.000&0.383&0.517&0.478&0.362&0.319&0.393&0.259&0.411 \\
20& 0.000&0.710&0.347&0.376&0.350&0.273&0.262&0.262&0.550 \\
 B& $^{+0.574}_{-\infty}$  &0.343&0.343&0.416&0.520&0.167
			   &0.114&0.089&0.496 \\
\hline
\end{tabular}
\end{table*}

Finally, it is convenient to point out here that we have done
a $\chi^2$ test of the compatibility of the Monte Carlo simulated
samples. The results are shown in Table 4, where the probability
of an independent observer to find the realizations compatible 
is shown for each pair of realizations. As it can be seen, this
probability can be as low as 0.01, which is the same to say 
that the corresponding luminosity functions are completely 
incompatible, even if they have derived from the same set 
of input parameters and selection criteria. Obviously, the 
conclusion is that for a reasonable number of objects in the 
restricted sample, the white dwarf luminosity function is 
dominated by the selection criteria.
 
\subsection{A bayesian analysis of the simulated samples}
 
As previously stated, changing the initial seed of the random 
function generator the Monte Carlo code provides different 
independent realizations of the white dwarf luminosity function. 
All these realizations are ``a priori'' equally good. Besides, 
since the number of objects that is used to compute the white 
dwarf luminosity function is relatively small, large deviations 
are expected, especially at relatively high luminosities for which 
the cooling timescales are short. This, in turn, results in very
probable underestimates of the associated uncertainties, especially
at luminosities larger than $\log(L/L_{\sun})\ga -3.0$. Consequently 
we have used bayesian statistical methods (Press 1996) to obtain a 
realistic estimation of the errors involved and the most probable
value of the density of white dwarfs for each luminosity bin. 

The problem can be stated as follows: for a given luminosity, $L$, we 
want to know the most probable value of the white dwarf luminosity 
function, $N$, given a set of $N_{\rm i}$ simulations assuming that 
all simulations are equally good. To compute $N$ one must maximize 
the probability distribution
\begin{equation}
P(N/N_{\rm i})\propto \prod_i \frac{1}{2}\Big(P_{\rm G_i}+P_{\rm B_i}
\Big)
\end{equation}
where $P_{\rm G}$ and $P_{\rm B}$ are the probability of being a good 
and a bad simulation, respectively. We can calculate them following 
closely Press (1996):
\begin{eqnarray}
P_{\rm G_i}&=&\exp\bigg[-\frac{(N_{\rm i}-N)^2}{2\sigma_{\rm i}^2}
\bigg]\nonumber
\\
&&\\
P_{\rm B_i}&=&\exp\bigg[-\frac{(N_{\rm i}-N)^2}{2S^2}\bigg]\nonumber
\end{eqnarray}
where $\sigma_{\rm i}$ is the error bar of each bin of the luminosity 
function and $S$ is a large but finite number characterizing the 
maximum expected deviation in $N_{\rm i}$. We recall that the 
contribution to the error of each white dwarf is equal to the 
inverse of its maximum volume squared.

The results are shown in Figure 11, where the probability 
distributions corresponding to each luminosity bin, computed 
with the previous method, are displayed. The logarithm of the 
luminosity of each bin in solar units is shown in the upper 
right corner of each panel. All the probability distributions, 
except that of the brighter luminosity bin, have a Gaussian 
profile. This is a direct consequence of the poor statistical 
significance of the first bin. In order to produce these 
probability distributions 20 independent realizations of the 
simulated samples were used. This is a reasonable number: increasing 
the total number of simulations does not introduces substantial 
improvements in the statistical significance of the first bin,
which is the less significant. From these probability distributions 
a better estimate of the statistical noise can be obtained. We have 
estimated the resulting error bars by assuming a conservative 95\% 
confidence level (approximately $2\sigma$). In Table 5 we show 
the computed deviations for each of the twenty realizations of the 
Monte Carlo simulated white dwarf luminosity functions and the 
most probable error bars computed at the 95\% confidence level.
The error bars obtained from a bayesian analysis of the twenty
Monte Carlo simulations compare favourably, roughly speaking, 
with those of each individual Monte Carlo simulation. However,
for samples where the total number of white dwarfs is smaller 
than 200 (the simulations presented here) the errors for each
of the luminosity bins are severe underestimates of the real 
errors, especially at low luminosities. 

\begin{figure}
\vspace{6.5cm}
\includegraphics{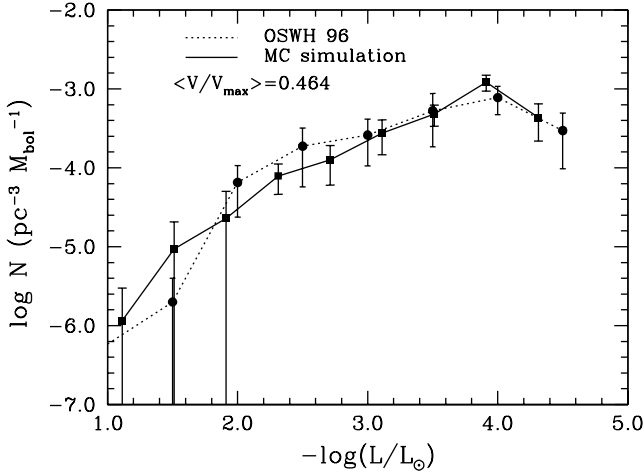}
\caption{Bayesian luminosity function.}
\end{figure}

In Figure 12 the most probable white dwarf luminosity function 
--- hereinafter bayesian white dwarf luminosity function --- with 
its corresponding error bars is shown, obtained by maximizing the 
probability distributions of figure 11. Except for moderately
high luminosities --- i.e. for luminosities larger than $\log(L/L_
{\sun})=-2.0$ --- where the effects of the spatial inhomogeinities
are most obvious the agreement between the observational luminosity
function and the bayesian luminosity function is excellent. Moreover, 
for the bayesian white dwarf luminosity we have computed a synthetic 
value of $\langle V/V_{\rm max}\rangle$ as an average of the 
corresponding values for each of the twenty realizations with
the weights given by the probability of each realization obtained
from the probability distributions of figure 11. We have obtained 
a value of $\langle V/V_{\rm max}\rangle=0.464$ which remains close 
to the canonical value of $\langle V/V_{\rm max}\rangle=0.5$, valid 
for an homogenous and complete sample.

\begin{figure}
\centering
\vspace{15cm}
\includegraphics{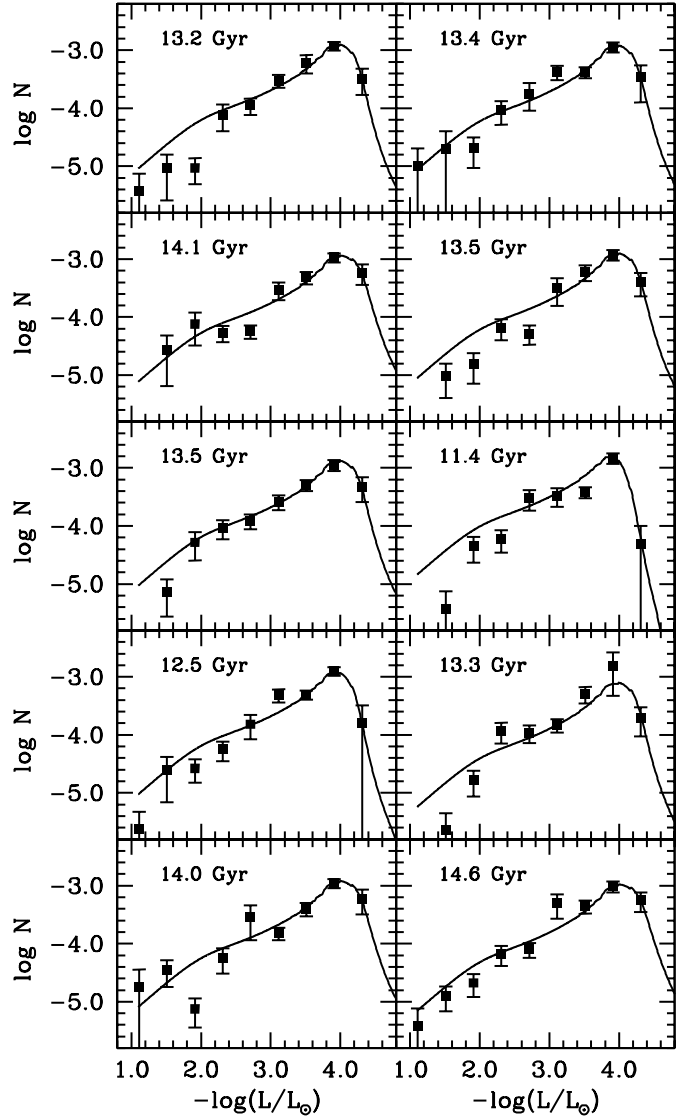}
\caption{Panel showing different simulated luminosity functions (filled 
squares) and their corresponding fit using a standard method.}
\end{figure}

\subsection{The age of the disk}

Perhaps one of the most surprising results of the simulations 
presented here is the age of the disk itself. The value of 13 
Gyr adopted in this paper fits nicely the observational data
of Oswalt et al. (1996) as can be seen in figure 12. This a 
direct consequence of the adopted scale height law, since using 
the same set of cooling sequences and a conventional approach to 
compute the white dwarf luminosity function with a constant 
volumetric star formation rate, Salaris et al. (1997) derived 
an age for the solar neighborhood of 11 Gyr when the effect of 
phase separation upon crystallization was taken into account and 
of 10 Gyr when phase separation was neglected. Thus the ultimate 
reason of the increase in the adopted age of the solar neighborhood 
is not due to the details of the adopted cooling sequences. Instead, 
this increase can be easily explained in terms of the model of 
galactic evolution. We recall that the white dwarf luminosity 
function measures the number of white dwarfs per {\sl cubic 
parsec} and unit bolometric magnitude. Therefore in order to 
evaluate it the volumetric star formation rate is required. In 
our case we can define the effective star formation rate per cubic 
parsec as $\psi_{\rm eff}(t)\approx\psi(t)/H_{\rm p}(t)$. With the 
laws adopted here for $\psi(t)$ and $H_{\rm p}(t)$ it is easy to 
verify that the effective star formation rate only becomes significant 
after $\sim\,2$ Gyr (Isern et al. 1995a,b). 

\subsection{Statistical uncertainties in the derived age of the disk}

The easiest and more straightforward way to assess the statistical 
errors associated with the measurement of the age of the solar 
neighborhood is trying to reproduce the standard procedure. That 
is, we have fitted the position of the ``observational'' cut-off
of each of the Monte Carlo realizations with a standard method 
(Hernanz et al. 1994) to compute the white dwarf luminosity
function using {\sl exactly} the same inputs adopted to simulate 
the Monte Carlo realizations, except, of course, the age of the 
disk, which is the only free parameter. The results are shown
in Figure 13 for ten of the twenty realizations. As is usual 
with real observational luminosity functions the theoretical 
white dwarf luminosity functions were normalized to the bin
with minimum error bars. The derived ages of the disk for each
one of the realizations are shown in the upper left corner of 
the corresponding panel. As it can be seen, there is a clear bias:
the derived ages of the disk are {\sl systematically} larger
than the input age of the Monte Carlo simulator by about half a
Gyr. This a direct consequence of the binning procedure, since
we are grouping white dwarfs belonging to the maximum of the 
white dwarf luminosity function in the lowest luminosity bin,
and can be avoided by using the cumulative white dwarf luminosity
function, which minimizes the effects of the binning procedure.

\begin{figure}
\vspace{6.5cm}
\includegraphics{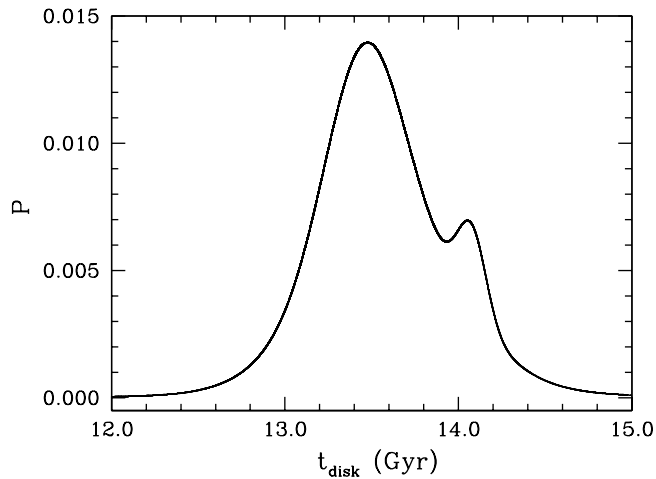}
\caption{Bayesian analysis of the derived age of the disk.}
\end{figure}

We have used the bayesian inference techniques described in \S 
4.3 to assign a purely statistical error to our age estimates. 
In order to do this we need to know a formal uncertainty for 
each one of the independent realizations. Since the value of
$\langle V/V_{\rm max}\rangle$ is a good measure of the 
overall quality of the sample (despite the fact that the 
samples are inhomogenous) we have adopted $\sigma_{\rm i}=
2\,(0.5-\langle V/V_{\rm max}\rangle)\,t_{\rm disk}$. The 
correspoding probability distribution is shown in Figure 14, 
which leads to a most probable age of the disk of $t_{\rm 
disk}=13.5\,\pm\,0.8$~Gyr at the 95\% confidence level 
($2\sigma$) which has to be compared with the adopted 
input age in our Monte Carlo simulation which was taken to
be 13~Gyr. This uncertainty is in good agreement with the 
results of Wood \& Oswalt (1998). If one relaxes the confidence 
level to $1\sigma$ the associated error bar is $\pm\,0.4$~Gyr. 
It is nonetheless important to realize that there is a systematic 
increase in the inferred disk ages of $\sim\,5$\%.

\section{Conclusions}

We have built a Monte Carlo code which has allowed us to reproduce
quite accurately the process of building the white dwarf luminosity
function from the stage of object selection and data binning. Our 
Monte Carlo simulation includes a model of galactic evolution based 
on well established grounds. The simulated samples obtained with 
our code reproduce very well the overall kinematical properties
of the observed white dwarf population. We have explored as well 
the temporal evolution of such properties and we have found that 
there is a fair agreement between the observed distributions and 
the simulated ones. However, we have shown that observed sample
has very few old white dwarfs and this has constrained our 
kinematical analysis to the most recent 3.7 Gyr, which can be 
considered as reasonably secure. Nevertheless, we have shown that 
the precise shape of the temporal distributions encode a wealth
of information and, therefore, more detailed analysis should be 
undertaken with improved observational samples. We have also 
extended our kinematical analysis by comparing the averaged
properties of both the observational white dwarf sample and 
the Monte Carlo simulated sample with a sample of old main 
sequence F and G stars. The results of this comparison lead 
to the conclusion that our model of galactic evolution is
fully compatible with the properties of the observed samples. 

Using our synthetic populations we have assessed the completeness 
of the samples used to derive the white dwarf luminosity function 
and studied their spatial distribution, and given a set of selection 
criteria consistent with the observational procedures, we have built 
several independent realizations of the white dwarf luminosity function 
and compared them to the observational luminosity function of Oswalt 
et al. (1996). Our results regarding the white dwarf luminosity 
function can be summarized as follows:

\begin{enumerate} 

\item Given the selection criteria adopted by Oswalt et al. 
(1996), our Monte Carlo simulation strongly suggest that the 
observational samples are complete up to $17^{\rm mag}$
and that the primary selection criterion at low luminosities 
is the proper motion one, in agreement with Wood \& Oswalt
(1998).

\item The Monte Carlo simulated white dwarf luminosity functions 
present an excellent agreement with the observational data. 

\item The effects of a scale height law are not negligible at all 
in the spatial distribution of the samples, especially at moderately 
large luminosities were they are more prominent. A scale height of 
roughly 130 pc is derived from the Monte Carlo simulations for the 
objects used in building the white dwarf luminosity function. However, 
we have established without any doubt that the effects of a scale 
height law should be more apparent in the cumulative distribution 
of proper motions. Although the effects of the scale height law 
on the tail of the white dwarf luminosity function seem to be 
negligible at first glance, a detailed analysis reveals that this 
inflation effect increases the derived ages of the disk by a 
considerable amount, which can be typically 2~Gyr. 

\item By using bayesian inference techniques we have been able 
to establish that the current procedure to assign the observational 
error bars to the white dwarf luminosity function is reasonably for 
a sample of 200 stars. 

\item Finally, the statistical uncertainty in the age of the disk 
derived from a bayesian analysis is roughly 1~Gyr, in agreement 
with Wood \& Oswalt (1998), and we have determined that there is 
a systematic trend due to the binning procedure which increases 
the disk ages inferred from the observational luminosity function 
by roughly a 5\%.

\end{enumerate} 

Nevertheless, a good deal of work remains to be done. Future 
improvements may include a more detailed analysis of the 
kinematical properties of the sample of old white dwarfs. For 
this purpose, it would be very useful to have more reliable 
observational samples. In our case this means not only complete 
samples but also more accurate mass determinations. It would be
also convenient to analyze the three dimensional motion of the 
white dwarf population but, for the moment, this seems to be 
unavoidable for the faintest white dwarfs due to the absence of
spectral features. Also of interest is to study the contamination 
of the input samples used in the process of building the white 
dwarf luminosity function with white dwarfs belonging to the 
galactic halo. This, in principle, cannot be discarded since 
halo members are selected on the basis of high proper motion, 
which is the dominant selection criterion at the dim end of 
the disk white dwarf luminosity function. A detailed statistical 
analysis of the cumulative counts of white dwarfs still remains 
to be done, instead of using the differential space density. 
Last but not least, the tests proposed in this paper could be 
applied to a real sample, thus providing us with very useful 
hints about the structure and evolution of our Galaxy.

\vspace{1 cm}

\noindent
{\sl Acknowledgements} This work has been supported by DGICYT grants 
		       PB94--0111 and PB94--0827-C02-02, and by the 
		       CIRIT grant GRQ94--8001.


\begin{thebibliography}{}

\bibitem{AB80}
Avni, Y., Bahcall, J.N., 1980, \apj, 235, 694
\bibitem{B92}
Bergeron, P., Saffer, R.A., Liebert, J., 1992, \apj, 394, 228
\bibitem{B95}
Bergeron, P., Wesemael, F., Beauchamp, A., 1995, \pasp, 107, 1047
\bibitem{B93}
Bravo, E., Isern, J., Canal, R., 1993, \aap, 270, 288
\bibitem{C90}
Carney, B.W., Latham, D.W., Laird, J.B., 1990, \aj, 99, 572
\bibitem{D97}
Dehnen, W., Binney, J.J., 1997, {\sl astro-ph/9710077} 
\bibitem{DP94}
D\'\i az-Pinto, A., Garc\'\i a-Berro, E., Hernanz, M., Isern,
J., Mochkovitch, R., 1994, \aap, 282, 86
\bibitem{E93}
Edvardsson, B., Andersen, J., Gustafsson, B., Lambert, D.L., Nissen, 
P.E., Tomkin, J., 1993, \aap, 275, 101
\bibitem{F76}
Felten, J.E., 1976, \aaps, 207, 700
\bibitem{F86}
Fleming, T.A., Liebert, J., Green, R.F., 1986, \apj, 308, 176
\bibitem{F96}
Flynn, C., Sommer-Larsen, J., Christensen, P.R., 1996, \mnras, 281,
1027
\bibitem{GB88}
Garc\'\i a-Berro, E., Hernanz, M., Mochkovitch, R., Isern, J.,
1988, \aap, 193, 141
\bibitem{GBT96}
Garc\'\i a-Berro, E., Torres, S., 1997, in White Dwarfs, ed. J.
Isern, M. Hernanz \& E. Garc\'\i a-Berro (Kluwer Academic Publishers),
97
\bibitem{G80}
Green, R.F., 1980, \apj, 238, 685
\bibitem{H94}
Hernanz, M., Garc\'\i a-Berro, E., Isern, J., Mochkovitch, R.,
Segretain, L., Chabrier, G., 1994, \apj, 434, 652
\bibitem{I89}
Iben, I., \& Laughlin, G., 1989, \apj, 341, 312
\bibitem{I95a}
Isern, J., Garc\'{\i}a-Berro, E., Hernanz, M., Mochkovitch, R., 
Burkert, A., 1995a, in White Dwarfs, ed. D. Koester \& K. Werner 
(Springer Verlag), 19
\bibitem{I95b}
Isern, J., Garc\'{\i}a-Berro, E., Hernanz, M., Mochkovitch, R., 
Burkert, A., 1995b, in The Formation of the Milky Way, ed. 
E.J. Alfaro \& A.J. Delgado (Cambridge University Press), 19
\bibitem{J90}
James, F., 1990, Comput. Phys. Commun., 60, 329
\bibitem{LRB88}
Leggett, S.K., Ruiz, M.T., Bergeron, P., 1998, \apj, in press
\bibitem{LDM88}
Liebert, J., Dahn, C.C., Monet, D.G., 1988, \apj, 332, 891
\bibitem{LDM89}
Liebert, J., Dahn, C.C., Monet, D.G., 1989, in White Dwarfs, ed.
G. Wegner (Springer Verlag), 15
\bibitem{M81}
Mihalas, D., Binney, J., 1981, Galactic Astronomy (W.H. Freeman \&
Co.)
\bibitem{N98}
Ng, Y.K., Bertelli, G., 1998, \aap, 329, 943
\bibitem{N90}
Noh, H.-R., Scalo, J., 1990, \apj, 352, 605 
\bibitem{O65}
Ogorodnikov, K.F., 1965, Dynamics of Stellar Systems (Pergamon Press)
\bibitem{OS95}
Oswalt, T.D., Smith, J.A., 1995, in White Dwarfs, ed. D. Koester \& K. 
Werner (Springer Verlag), 24
\bibitem{OS96}
Oswalt, T.D., Smith, J.A., Wood, M.A., Hintzen, P., 1996, \nat, 382, 692
\bibitem{P96}
Press, W.H., 1996, in Unsolved problems in Astrophysics (Princenton
University Press)
\bibitem{P86}
Press, W.H., Flannery, B.P., Teukolsky, S.A., Vetterling, W.T., 1986,
Numerical Recipes (Cambridge University Press)
\bibitem{Q97}
Qin, Y.P., \& Xie, G.Z., 1997, \apj, 486, 100
\bibitem{S97}
Salaris, M., Dom\'{\i}nguez, I., Garc\'{\i}a-Berro, E., Hernanz, M., 
Isern, J., Mochkovitch, R., 1997, \apj, 486, 413
\bibitem{S61}
Salpeter, E.E., 1961, \apj, 134, 669
\bibitem{S98}
Scalo, J., 1998, in The Stellar Initial Mass Function, eds. G. Gilmore,
I. Parry \& S. Ryan (PASP Conference Series), in press
\bibitem{Sc68}
Schmidt, M., 1968, \apj, 151, 393
\bibitem{S88}
Sion E.M., Fritz, M.L., McMullin, J.P., Lallo, M.D., 1988, \aj, 96, 251
\bibitem{S77}
Sion E.M., Liebert, J., 1977, \apj, 213, 468
\bibitem{T93}
Tinney, C.G., Reid, I.N., Mould, J.R., 1993, \apj, 414, 254
\bibitem{W87}
Winget, D.E., Hansen, C.J., Liebert, J., Van Horn, H.M., Fontaine, G., 
Nather, R., Kepler, S.O., Lamb, D.K., 1987, \apj, 315, L77
\bibitem{W92}
Wood, M.A., 1992, \apj, 386, 539
\bibitem{W97}
Wood, M.A., 1997, in White Dwarfs, ed. J. Isern, M. Hernanz \& E. 
Garc\'\i a-Berro (Kluwer Academic Publishers), 105
\bibitem{WO98}
Wood, M.A., Oswalt, T.D., 1998, \apj, in press

\end{thebibliography}
\end{document}